\documentclass[journal]{IEEEtran}

\IEEEoverridecommandlockouts                              % This command is only
\usepackage{amssymb}
\usepackage{amsmath}
\usepackage{bm}
\usepackage{mathrsfs}
\usepackage{comment} 
\usepackage{algorithmic,algorithm}
\usepackage{shuffle}
\usepackage{graphicx} % for pdf, bitmapped graphics files
\usepackage{caption}
\usepackage{lipsum}
\usepackage{color}
\usepackage{enumerate}
\usepackage{amsfonts}
\usepackage{subfigure} 
\usepackage{array}
\usepackage{enumerate}
\usepackage{setspace}
\usepackage{multirow}
\DeclareSymbolFont{symbolsC}{U}{pxsyc}{m}{n}
\DeclareMathSymbol{\coloneqq}{\mathrel}{symbolsC}{"42}

\begin{document}

\title{\LARGE \bf
Synthesis of Covert Sensor Attacks in Networked Discrete-Event Systems with Non-FIFO Channels} %Partially Observed  via Projections of Holons
\author{Ruochen Tai, Liyong Lin, Yuting Zhu and Rong Su
%\thanks{
%This work was partially supported by the National Natural Science Foundation of China under Grant
%Nos. 61374068, 61472295, and 61672400, the Recruitment Program of Global Experts, and the Science
%and Technology Development Fund, MSAR, under Grant Nos. 078/2015/A3 and 106/20156/A3
%(Corresponding author: Z. Li).

%D.~Wang is with the School of Electro-Mechanical Engineering, Xidian University, Xi'an 710071, China (e-mail: wdeguang1991@163.com).
%
%L.~Lin is with the Systems Control Group, Department of Electrical
%and Computer Engineering, University of Toronto, Toronto, ON M5S 3G4
%Canada (e-mail: liyong.lin@utoronto.ca).
%
%Z. Li is with the Institute of Systems Engineering, Macau University of Science and Technology, Taipa,
%Macau and also with the Key Laboratory of Electronic Equipment Structure Design, Ministry of
%Education, School of Electro-Mechanical Engineering, Xidian University, Xi'an 710071, China (e-mail: zhwli@xidian.edu.cn).
%
%W.~M.~Wonham is with the Systems Control Group, Department of Electrical
%and Computer Engineering, University of Toronto, Toronto, ON M5S 3G4,
%Canada (e-mail: wonham@control.utoronto.ca).}
%}
\thanks{The research of the project was supported by Ministry of Education, Singapore, under grant AcRF TIER 1-2018-T1-001-245 (RG 91/18).

The authors are affliated with Nanyang Technological University, Singapore. (Email: ruochen001@e.ntu.edu.sg; liyong.lin@ntu.edu.sg; yuting002@e.ntu.edu.sg; rsu@ntu.edu.sg).
(\emph{Corresponding author: Liyong Lin})}
}
\maketitle

%\begin{comment}
\begin{abstract}
In this paper, we investigate the covert sensor attack synthesis problem in the framework of supervisory control of networked discrete-event systems (DES), where the observation channel and the control channel are assumed to be non-FIFO and have bounded network delays. We focus on the class of sensor attacks satisfying the following properties: 1) the attacker might not have the same observation capability as the networked supervisor; 2) the attacker aims to remain covert, i.e., hide its presence against the networked monitor; 3) the attacker could insert, delete, or replace compromised observable events; 4) it performs bounded sensor attacks, i.e., the length of each string output of the sensor attacker is upper bounded by a given constant. The solution methodology proposed in this work is to solve the covert sensor attack synthesis problem for networked DES by modeling it as the well studied Ramadge-Wonham supervisor synthesis problem, and the constructions work for both the damage-reachable attacks and the damage-nonblocking attacks.
In particular, we show the supremal covert sensor attack exists in the networked setup and can be effectively computed by using the normality property based synthesis approach. 
\end{abstract}


%{\color{red} safety and liveness, mu caluclus, buchi automaton}

%{\color{red} title may need change}

\section{Introduction}
\label{intro}
As an integration of cyber information and physical world, cyber-physical systems (CPS) have been playing a significant role in the modern society due to the precise control, remote collaboration and autonomous functions. The realization of these powerful features heavily relies on the network system (cyber part), which might be compromised and exploited to cause irreparable damage by malicious attacks. 
Recently, the security issue of CPS has drawn a lot of attention from both the computer science community and the systems control community. Quite a few works have been devoted to the cyber security issues related to control, optimization, and computation \cite{cardenas2008secure}-\cite{zhang2015optimal}. 
%\cite{cardenas2008secure,fawzi2014secure,teixeira2012attack,mo2011cyber,zhang2015optimal}
As a class of attacks which is able to remain hidden until the damage is caused to the system, covert attacks
could avoid being detected, successfully bring about system breakdown, and impose greater threats on the secure operation.
It is thus of importance to investigate covert attacks in CPS. 

Based on the locations of adversial actions, covert attacks could be divided into three categories: 1) sensor attacks (observation channel); 2) actuator attack (control channel); 3) sensor-actuator attack (observation and control channel). In this work, we shall study the covert sensor attack synthesis problem in CPS which is modeled as networked DES. 
For sensor attacks, \cite{su2018supervisor}-\cite{meira2019synthesis}  %\cite{su2018supervisor,meira2020synthesis,meira2019synthesis} 
consider the attacks that can alter the sensor readings. In these papers, it is assumed that the attack and the supervisor have the same observation ability. 
%2) each time the attack observes one event, it would output one (altered) sensor reading based on its attack strategy. 
In \cite{su2018supervisor}, the attack is modeled as a finite state transducer and needs to guarantee damage-infliction to the system no matter which trajectory of the closed-loop system is executed. \cite{su2018supervisor} proposes an approach to synthesize the supremal covert sensor attack under a normality condition, based on which a resilient supervisor can be synthesized. \cite{meira2020synthesis} builds a game arena and employs a game-theoretic technique to synthesize maximally permissive covert sensor attacks. 
In \cite{meira2019synthesis}, the plant is modeled as a probabilistic automaton and the attack synthesis problem is transformed into an optimization problem which is solved by the approaches used in stochastic graph-games.

%In the context of DES, actuator attack could modify the control command issued by the supervisor. 
%\cite{lin2019synthesis} presents a new problem formulation for the synthesis of covert actuator attacks, assuming that the attack could eavesdrop the information sent by the plant and supervisor. 
%In \cite{lin2020synthesis}, a generalized treatment of actuator attack synthesis is proposed, where the synthesis problem is transformed into the Ramadge-Wonham supervisor synthesis problem, a well-studied control problem. To make the supervisor resilient against actuator attacks, \cite{zhu2019supervisor} proposes an obfuscation approach to preserve the behavior of the original closed-loop system, where the
%boolean satisfiability problem (SAT) solver \cite{grumberg2004memory} is adopted to synthesize the obfuscated supervisor. 
%In \cite{carvalho2016detection}, the notion of AE (actuator enablement)-safe controllability is proposed, serving as the base to validate whether the attack can be identified by a diagnoser before an unsafe state is reached. {\color{red} need to mention actuator attack?}

For sensor-actuator attacks, \cite{carvalho2018detection}-\cite{lima2018detectable} %\cite{carvalho2018detection,lima2017security,lima2018detectable} 
focus on the detection of attacks and adopt the strategy of disabling all controllable events after the attack is detected. 
GF (general form)-safe controllability is defined in \cite{carvalho2018detection} to formalize whether the attack can be identified by a diagnoser before an unsafe state is reached. 
As an extension of \cite{lima2017security}, \cite{lima2018detectable} presents the notions of detectable network attack security (DNA-security) and the undetectable network attack security (UNA-security). These security notions proposed in \cite{carvalho2018detection}-\cite{lima2018detectable} could be checked via algorithms derived from the ones for the diagnosability check.
\cite{Lin2019WODES}-\cite{wang2019attack}
%\cite{Lin2019WODES,Liyong2020attack,lin2019towards,lin2020bounded,wang2019attack} 
focus on the synthesis issue.
To synthesize covert sensor-actuator attacks, \cite{Lin2019WODES} and \cite{Liyong2020attack} transform it into the Ramadge-Wonham supervisor synthesis problem, where the attacks can eavesdrop the information in the observation channel and the control channel. 
 In \cite{lin2019towards} and \cite{lin2020bounded}, the problem of bounded synthesis of resilient supervisors against sensor-actuator attacks is reduced to the Quantified Boolean Formula (QBF) problem, which could be solved by the QBF solver or with repeated calls to the propositional satisfiability (SAT) solver. Resilient control can also be achieved by adding an artificial secure channel for the control commands, as studied in \cite{wang2019attack}. 

As we have introduced above, lots of fruitful works focus on security issues in the context of DES. However, the shared communication network, an indispensable ingredient in CPS, would unavoidably induce channel delays that cannot be neglected; thus, networked DES is more suitable for capturing the properties of CPS. In networked DES, more difficulties and opportunities are presented for the synthesis of covert sensor attacks, which renders it more complex than the counterpart in the non-networked setup:  
%{\color{red} may need to reorganize: 1. what makes it more difficult for the attack: a) supervisor is already resilient against disordering, b) non-FIFO channel may expose the attack. 2. what makes it easier for the attack: networked monitor harder to detect. }
\begin{itemize}
\setlength{\itemsep}{3pt}
\setlength{\parsep}{0pt}
\setlength{\parskip}{0pt}
    \item Difficulties: To address the non-FIFO channels in networked DES in the absence of attacks, the designed networked supervisor should be resilient enough such that any disordered event sequence caused by channel delays can already be handled. Thus, some attack sequences that could have been used by the attacker to cause damage in the non-networked setup would not be effective anymore for the attacker in networked DES if these attack sequences are included in the above-mentioned disordered event sequences, which makes it harder for the attacker to cause damage-infliction. 
    %Constrained by non-FIFO channels, the sensor attacks in networked DES are in dilemma to alter sensor readings as the random channel delays could make it highly possible that such inconsistent information flow popped out from the observation channel would expose its existence to the networked monitor, which imposes great difficulties for the synthesis of covert sensor attacks in networked setup.
    \item Opportunities: Weakness also follows in the above-mentioned resilient networked supervisor since now it is ambiguous for the networked monitor to infer whether the disordered event sequences are caused by channel delays or information tampering by attacks, making it easier for the attack to remain covert. It is such weakness that the sensor attacker could take advantage of to implement attacks successfully and not detected by the networked monitor.
    %\item Since the above-mentioned opportunities and challenges coexist, attacks need to weigh the pros and cons between them such that, on one hand, the altered sensor readings could deceive the networked supervisor into issuing inappropriate control decisions, on the other hand, such fake information will not ruin the covertness property. 
\end{itemize}
Thus, it is of great significance and practical application values to investigate covert sensor attacks for CPS modeled as networked DES.
There are lots of works dedicated to the modeling and synthesis problem in networked DES. \cite{balemi1994input} proposes an input/output semantics to deal with communication delays. \cite{park2007supervisory} studies the existence of nonblocking supervisors in networked DES. \cite{Lin:14} and \cite{zhusupervisor} adopt the setup that channel delays are quantified by the number of executed events. In \cite{Lin:14}, 
%observation channel is non-FIFO and control channel is FIFO, then 
the model of the two channels are captured by mappings, based on which network controllability and observability are defined. In \cite{zhusupervisor}, a new networked control framework is proposed, where the observation channel and control channel are modeled as two finite state automata, based on which the synthesis of networked supervisors is transformed into a partial observation supervisor synthesis problem.
In \cite{rashidinejad2018supervisory}, 
%the observation channel is non-FIFO and the control channel is FIFO. 
tick event is adopted to quantify channel delays, and \cite{rashidinejad2018supervisory} adopts the idea of synthesizing a predictive supervisor to obtain a networked supervisor. The asynchronous enablement, execution, and observation of an event caused by communication delays can also be modeled by the asynchronous plant proposed in \cite{Rashidinejad2019asynchronous}.  Asynchronous supervisors satisfying controllability and nonblockingness properties are synthesized in \cite{Rashidinejad2019asynchronous}.

In networked DES, it is possible that plant $G$ might receive control commands containing no executable events w.r.t. the current state of $G$. The discarding strategy for such control commands adopted in previous works might lose ones useful for the later event executions at plant $G$ and cause undesired blockings.
Thus, to deal with this issue, in this work, we propose a new mechanism for the plant $G$ where the received control commands by $G$ would always be stored in the memory for a predefined time interval and $G$ would always fetch some stored control command containing executable events from the memory. By this way, our new mechanism of plant $G$ could alleviate the negative impacts brought about by the discarding strategy.

Based on the above new mechanism of plant $G$, in this paper, we study the synthesis problem of covert sensor attacks in networked DES with non-FIFO channels, which is a more realistic model for real systems. To the best of our knowledge, this is the first time to investigate the attack synthesis problem in networked DES. Before, the first and only work to study the attack in networked DES is \cite{Li2020detectandpreventofattck}, where the actuator enablement (AE) attack is taken into consideration and an algorithm to verify the proposed AE-safe controllability is developed. The difference between \cite{Li2020detectandpreventofattck} and our work is that \cite{Li2020detectandpreventofattck} focuses on the attack detection problem while our work solves the attack synthesis problem. Another work related to the security issue in networked DES is \cite{YinOpacityNetwork} while it investigates the opacity enforcement instead of the covert attack problem.
In this work, we adopt tick event to measure the passage of time and study the sensor attack synthesis problem in a general setup: 1) the sensor attack and the networked supervisor might have different observation capabilities; 2) the observation channel and control channel are both non-FIFO; 3) different event executions at plant $G$ might take up different numbers of tick events. The networked DES under sensor attack is composed of six components shown in Fig. \ref{fig:Networked supervisory control architecture under sensor attacks}: 1) plant with command execution and storage; 2) sensor attack; 3) observation channel; 4) networked supervisor; 5) networked monitor (used for detecting the attack); 6) control channel. 
%For the component, plant with command execution and storage, based on our new mechanism of plant $G$, it is a combination of the command storage, command execution, and state transition of plant $G$. 
The sensor attack studied in this work could implement insertion, deletion, and replacement attacks. Finite state automaton is adopted to model the dynamics of each component, base on which the connection between the sensor attack model and the observation channel model is established. 
%The closed-loop behavior of the networked system under sensor attack is generated by the synchronous product of the above-mentioned six component models. 
Then, we propose the methodology of modeling the covert sensor attack synthesis problem in networked DES with non-FIFO channels as the Ramadge-Wonham supervisory control problem, and explain how the supremal covert sensor attack can be computed based on the normality property. %Finally, we provide the sufficient and necessary condition of the existence of the covert attack for both the damage-nonblocking and damage-reachable goal.

This paper is organized as follows. In Section \ref{sec:Preliminaries}, we provide some basic notions which are needed in this work. In Section \ref{sec:Component Models for Networked DES under sensor attacks}, we introduce the formalization of components in networked DES under sensor attack. Section \ref{sec:Synthesis of Covert Sensor attacks for Networked DES} explains our method for solving the synthesis problem of covert sensor attacks for networked DES. An example is then given to illustrate the effectiveness of the proposed method in Section \ref{sec:example}. Finally, conclusions are drawn in Section \ref{sec:conclusions}.

\section{Preliminaries}
\label{sec:Preliminaries}
Given a finite alphabet $\Sigma$, let $\Sigma^{*}$ be the free monoid over $\Sigma$ with the empty string $\varepsilon$ being the unit element and the string concatenation being the monoid operation. For a string $s$, $|s|$ is defined as the length of $s$. Given two strings $s, t \in \Sigma^{*}$, we say $s$ is a prefix substring of $t$, written as $s \leq t$, if there exists $u \in \Sigma^{*}$ such that $su = t$, where $su$ denotes the concatenation of $s$ and $u$. A language $L \subseteq \Sigma^{*}$ is a set of strings. The prefix closure of $L$ is defined as $\overline{L} = \{u \in \Sigma^{*} \mid (\exists v \in L) \, u\leq v\}$. 
%If $L = \overline{L}$, then $L$ is \emph{prefix-closed}. The concatenation of two languages $L_{a}, L_{b} \subseteq \Sigma^{*}$ is defined as $L_{a}L_{b} = \{s_{a}s_{b} \in \Sigma^{*}|s_{a} \in L_{a} \wedge s_{b} \in L_{b}\}$.
The event set $\Sigma$ is partitioned into $\Sigma = \Sigma_{c} \dot{\cup} \Sigma_{uc} = \Sigma_{o} \dot{\cup} \Sigma_{uo}$, where $\Sigma_{c}$ (respectively, $\Sigma_{o}$) and $\Sigma_{uc}$ (respectively, $\Sigma_{uo}$) are defined as the sets of controllable (respectively, observable) and uncontrollable (respectively, unobservable) events, respectively.  As usual, $P_{o}: \Sigma^{*} \rightarrow \Sigma_{o}^{*}$ is the natural projection defined such that
\begin{enumerate}[(1)]
\item $P_{o}(\varepsilon) = \varepsilon$,
\item $(\forall \sigma \in \Sigma) \, P_{o}(\sigma)=
\left\{
\begin{array}{rcl}
\sigma       &      & {\sigma \in \Sigma_{o},}\\
\varepsilon  &      & {\rm otherwise,}
\end{array} \right.$
\item $(\forall s \in \Sigma^*, \sigma \in \Sigma) \, P_{o}(s\sigma) = P_{o}(s)P_{o}(\sigma)$.
\end{enumerate}

A finite state automaton $G$ over $\Sigma$ is given by a 5-tuple $(Q, \Sigma, \xi, q_{0}, Q_{m})$, where $Q$ is the state set, $\xi: Q \times \Sigma \rightarrow Q$ is the (partial) transition function, $q_{0} \in Q$ is the initial state, and $Q_{m}$ is the set of marker states. 
We write $\xi(q, \sigma)!$ to mean that $\xi(q, \sigma)$ is defined and also view $\xi \subseteq Q \times \Sigma \times Q$ as a relation. $En_{G}(q) = \{\sigma \in \Sigma|\xi(q, \sigma)!\}$.
$\xi$ is also extended to the (partial) transition function $\xi: Q \times \Sigma^{*} \rightarrow Q$ and the transition function $\xi: 2^{Q} \times \Sigma \rightarrow 2^{Q}$ \cite{wonham2015supervisory}, where the later is defined as: for any $Q^{'} \subseteq Q$ and any $\sigma \in \Sigma$, $\xi(Q^{'}, \sigma) = \{q^{'} \in Q|(\exists q \in Q^{'})q^{'} = \xi(q, \sigma)\}$. 
Let $L(G)$ and $L_{m}(G)$ denote the closed-behavior and marked behavior, respectively. When $Q_{m} = Q$, we shall also write $G = (Q, \Sigma, \xi, q_{0})$ for simplicity. 
The ``unobservable reach'' of the state $q \in Q$ under the subset of events $\Sigma^{'} \subseteq \Sigma$ is given by $UR_{G, \Sigma - \Sigma^{'}}(q) := \{q^{'} \in Q|[\exists s \in (\Sigma - \Sigma^{'})^{*}] \, q^{'} = \xi(q,s)\}$.
We shall abuse the notation and define the subset construction $P_{\Sigma^{'}}(G)$ to be the finite state automaton $(2^{Q}, \Sigma, \delta, UR_{G, \Sigma - \Sigma^{'}}(q_{0}))$ over $\Sigma$, where $UR_{G, \Sigma - \Sigma^{'}}(q_{0}) := \{q \in Q|[\exists s \in (\Sigma - \Sigma^{'})^{*}] \, q = \xi(q_{0},s)\}$ of $q_{0}$ \cite{wonham2015supervisory} is the initial state, and the (partial) transition function $\delta: 2^{Q} \times \Sigma \rightarrow 2^{Q}$ is defined as follows:
\begin{enumerate}[(1)]
    \item For any $\varnothing \neq Q^{'} \subseteq Q$ and any $\sigma \in \Sigma^{'}$, $\delta(Q^{'}, \sigma) = UR_{G, \Sigma - \Sigma^{'}}(\xi(Q^{'}, \sigma))$, where
    \[
    UR_{G, \Sigma - \Sigma^{'}}(Q^{''}) = \bigcup\limits_{q \in Q^{''}}UR_{G, \Sigma - \Sigma^{'}}(q)
    \]
    for any $Q^{''} \subseteq Q$ and $UR_{G, \Sigma - \Sigma^{'}}(q)$ is the unobservable reach of $q$;
    \item For any $\varnothing \neq Q^{'} \subseteq Q$ and any $\sigma \in \Sigma - \Sigma^{'}$, $\delta(Q^{'}, \sigma) = Q^{'}$.
\end{enumerate}
It is noteworthy that $P_{\Sigma^{'}}(G)$ is over $\Sigma$, instead of $\Sigma^{'}$, and there is no transition defined at the state $\varnothing \in 2^{Q}$.
%\vspace{-0.3cm}

A finite state automaton $G = (Q, \Sigma, \xi, q_{0}, Q_{m})$ is said to be non-blocking if every reachable state in $G$ can reach some marked state in $Q_{m}$ \cite{wonham2015supervisory}. As usual, for any two finite state automata $G_{1} = (Q_{1}, \Sigma_{1}, \xi_{1}, q_{1,0}, Q_{1,m})$ and $G_{2} = (Q_{2}, \Sigma_{2}, \xi_{2}, q_{2,0}, Q_{2,m})$, where $En_{G_{1}}(q) = \{\sigma|\xi_{1}(q, \sigma)!\}$ and $En_{G_{2}}(q) = \{\sigma|\xi_{2}(q, \sigma)!\}$, their synchronous product is denoted as $G_{1}||G_{2} := (Q_{1} \times Q_{2}, \Sigma_{1} \cup \Sigma_{2}, \zeta, (q_{1,0}, q_{2,0}), Q_{1,m} \times Q_{2,m})$, where the (partial) transition function $\zeta$ is defined as follows, for all $(q_{1}, q_{2}) \in Q_{1} \times Q_{2}$ and $\sigma \in \Sigma$:
\[
\begin{aligned}
& \zeta((q_{1}, q_{2}), \sigma) := \\ & \left\{
\begin{array}{lcl}
(\xi_{1}(q_{1}, \sigma), \xi_{2}(q_{2}, \sigma))  &      & {\rm if} \, {\sigma \in En_{G_{1}}(q_{1}) \cap En_{G_{2}}(q_{2}),}\\
(\xi_{1}(q_{1}, \sigma), q_{2})       &      & {\rm if} \, {\sigma \in En_{G_{1}}(q_{1}) \backslash \Sigma_{2},}\\
(q_{1}, \xi_{2}(q_{2}, \sigma))       &      & {\rm if} \, {\sigma \in En_{G_{2}}(q_{2}) \backslash \Sigma_{1},}\\
{\rm not \, defined}  &      & {\rm otherwise.}
\end{array} \right.
\end{aligned}
\]
\textbf{Notation.} Let $\mathbb{Z}$ denote the set of integers, $\mathbb{N}$ the set of nonnegative integers, and $\mathbb{N}^{+}$ the set of positive integers. 
Let $\Gamma = 2^{\Sigma_c}-\{\varnothing\}$ be the set of all control commands, deviating from the standard definition of $\Gamma$. In this work, each control command only contains the controllable events that are enabled and it is assumed that when no control command is received by plant $G$, then only uncontrollable events could be executed. 
%and $(\Gamma^{*})^{\varnothing} = \Gamma^{*} \cup \{\varnothing\}$. 
%We will often use $\varnothing$ to physically denote ``empty message'' and
$tick$ denotes the tick event commonly used in the timed DES literature \cite{brandin1994supervisory}.
Let $[m:n] := \{m,m+1,\cdots,n\}$ ($m \in \mathbb{N}, n \in \mathbb{N}$) and $(m:n) := \{m+1,\cdots,n-1\}$ ($m \in \mathbb{N}, n \in \mathbb{N}$). By convention, when $m > n$, $[m:n] := \varnothing$, and when $m+1 > n-1$, $(m:n) := \varnothing$.
We use $\mathbb{N}^{S}$ to denote the set of all possible multisets whose underlying set is $S$. 
To avoid infinity, we shall restrict $\mathbb{N}^{S}$ by a finite parameter, say $\mathbb{N}^{S}(l)$, where $l$ denotes the bound on the cardinality of the multiset, counting the multiplicity. In Section \ref{subsec:Observation Channel} and \ref{subsec:Control Channel}, we shall adopt $\mathbb{N}^{S}(l)$ to denote the bounded channel messages, where $l$ denotes the maximum number of messages in a channel, which is always finite, considering bounded channel delays. 
For example, $\mathbb{N}^{\{1\}}(2) = \{\varnothing, \{1\}, \{1, 1\}\}$. 
We use $S^{in}$/$S^{out}$/$S^{\#}$ to denote a relabelled copy of $S$ with superscript ``$in$''/``$out$''/``$\#$'' attached to each element in $S$. Intuitively speaking, the superscript ``$in$'' denotes the entering of a message into the channel, ``$out$'' denotes the pop out of a message from the channel, and ``$\#$'' denotes the message tampering by the attacker. The specific meanings about the relabelled events will be introduced later in Section \ref{sec:Component Models for Networked DES under sensor attacks}.
%To avoid infinity, we shall restrict $\Sigma^{*}$ by a finite parameter $l$ and 
We define $\Sigma^{\leq l} = \{s \in \Sigma^{*}||s| \leq l\}$. For example, if $\Sigma = \{a\}$, then $\Sigma^{\leq 2} = \{\varepsilon, a, aa\}$.

\section{Component Models for Networked DES under Sensor Attack}
\label{sec:Component Models for Networked DES under sensor attacks}

%{\color{red} before introducing the technical content, need to motivate the reader what is interesting in this setup, what difficulty non-FIFO channels brings  to  the sensor attack, the monitor, and the resilient supervisor, for example, in some cases, the attack will succeed in the non-networked setup, using a simple example, but then fails or has reduces solution in the networked setup. The example can be given in example section}

\begin{figure}[htp]
\begin{center}
\includegraphics[height=7.1cm]{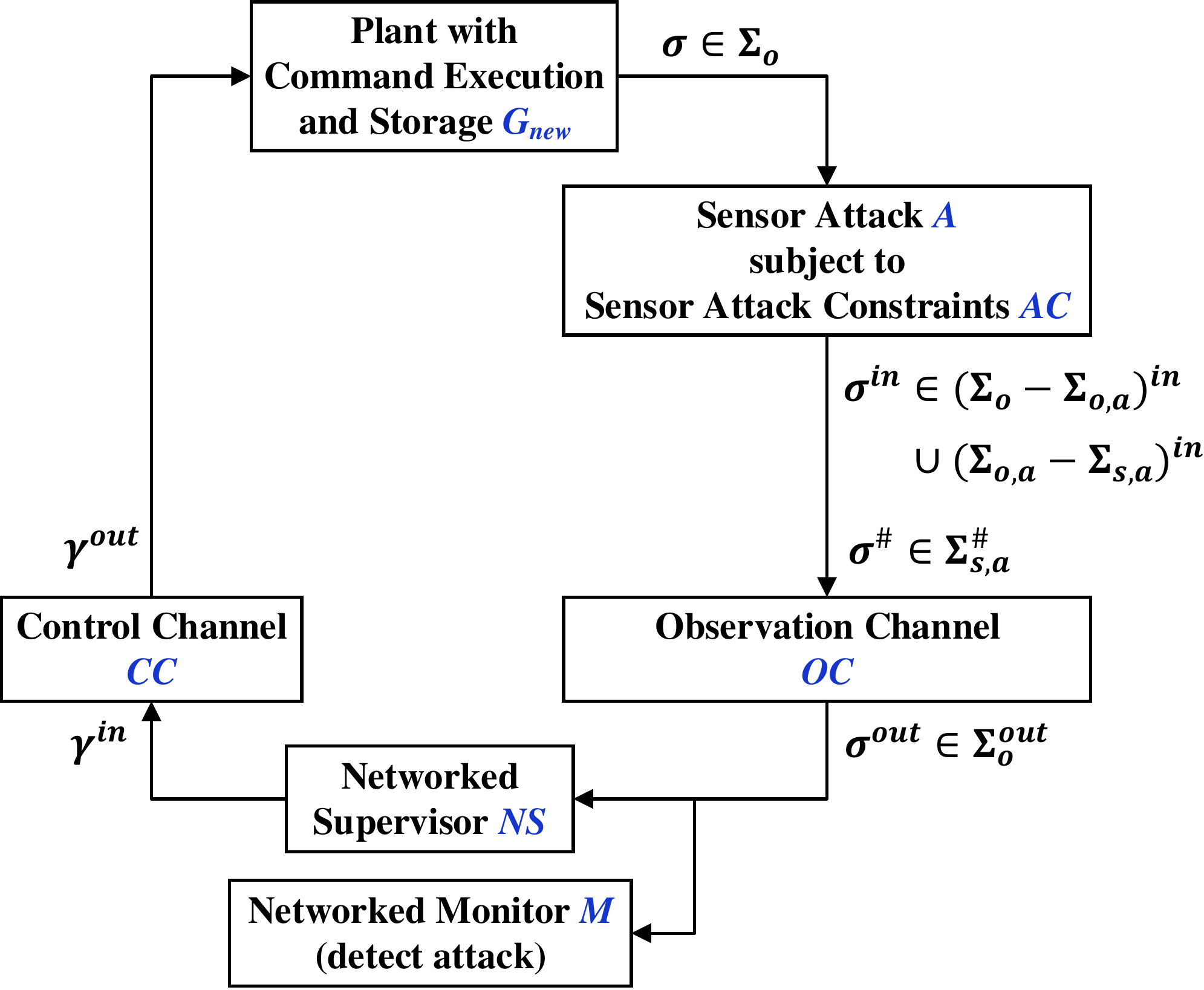}   
\caption{Networked supervisory control architecture under sensor attack}
\label{fig:Networked supervisory control architecture under sensor attacks}
\end{center}        
%{\color{red} later can add a picture, as "equ.png" drawn in the "WODES\_attack ppt" in the folder to explain why these modules are needed.}
\end{figure}

\vspace{-0.3cm}

The system architecture of networked DES under sensor attack is illustrated in Fig. \ref{fig:Networked supervisory control architecture under sensor attacks}, where the components are listed as follows:
\begin{itemize}
    \setlength{\itemsep}{3pt}
    \setlength{\parsep}{0pt}
    \setlength{\parskip}{0pt}
    \item Sensor attack (subject to sensor attack constraints).
    \item Observation channel.
    \item Control channel.
    \item Plant with command execution and storage. 
    \item Networked supervisor.
    \item Networked monitor (serves to detect the attack).
\end{itemize}

%{\color{red} may need to explain the challenge in the synthesis of sensor attack in this setup} {\color{blue}explained in Introduction}

In the following subsections, we shall explain how to model the above-mentioned six components and the specific meanings of the relabelled events presented in Fig. \ref{fig:Networked supervisory control architecture under sensor attacks}. 

\subsection{Sensor Attack}
\label{subsec:Sensor attack}

%\begin{figure}[htp]
%\begin{center}
%\includegraphics[height=3.4cm]{fig/Sensor_attack_Template_Normal.pdf}   
%\caption{Sensor attack template ($\Sigma_{s,a} \subseteq \Sigma_{o,a}$, $\sigma_{1}, \dots, \sigma_{U} \in \Sigma_{s,a}$)}
%\label{fig:Sensor attack template (normal)}
%\end{center}                                
%\end{figure}

In this work, the set of observable events for the sensor attacker is denoted as $\Sigma_{o,a} \subseteq \Sigma_{o}$, where $\Sigma_o$ denotes the set of observable events for the networked supervisor. Then, $\Sigma_{o} - \Sigma_{o,a}$ denotes the set of events that can be observed by the networked supervisor but cannot be observed by the attacker. The set of compromised observable events for the sensor attacker is denoted as $\Sigma_{s,a} \subseteq \Sigma_{o,a}$. We shall henceforth refer to $(\Sigma_{o,a}, \Sigma_{s,a})$ as an attack constraint.
The basic assumptions of the sensor attacker in this work are given as follows:
\begin{itemize}
    \item The sensor attacker is deployed at the entrance of the observation channel, as illustrated in Fig. \ref{fig:Networked supervisory control architecture under sensor attacks}.%\footnote{In practice, the attack needs to stay as far away from the networked supervisor as possible to prevent detection at the physical level. {\color{blue} Not sure if this is true as there can be compromise on the PLC controllers. The assumption that the sensor attack is deployed at the side of the plant is reasonable itself.}}
    \item The sensor attacker can only implement insertion, deletion, and  replacement attacks.
    \item When an attack is initiated for a specific observation, it will be completed before the next observation is generated by the plant $G$. Each time when the sensor attacker observes one event, the number of events that the attacker can simultaneously send into the observation channel is bounded by $U$, i.e., we consider bounded sensor attacks as in \cite{su2018supervisor}. 
    \item The sensor attack action (insertion, deletion, and replacement) initiated by the sensor attacker is instantaneous.
\end{itemize}

Next, we shall introduce two models that will be used in this work: 1) sensor attack constraints; 2) sensor attack, where the former one serves as a ``template'' to describe the attack capabilities and the latter one is the attack that we aim to synthesize. 

\textbf{Sensor Attack Constraints:} The sensor attack constraints is modeled as a finite state automaton $AC$, which simulates the finite state transducer model of the sensor attack of \cite{su2018supervisor}.

\[
AC = (Q_{ac}, \Sigma_{ac}, \xi_{ac}, q_{ac}^{init})
\]
\begin{itemize}
    \setlength{\itemsep}{3pt}
    \setlength{\parsep}{0pt}
    \setlength{\parskip}{0pt}
    \item $Q_{ac} = \{q_{o}^{\sigma}|\sigma \in \Sigma_{o,a} - \Sigma_{s,a}\} \cup 
    \{q_{uo}^{\sigma}|\sigma \in \Sigma_{o} - \Sigma_{o,a}\} \cup
    \{q_{n}|n \in [0: U]\} \cup
    \{q_{ac}^{init}\}$
    \item $\Sigma_{ac} = \Sigma \cup (\Sigma_{o} - \Sigma_{s,a})^{in} \cup \Sigma_{s,a}^{\#} \cup \Sigma_{o}^{out} \cup \Gamma^{in} \cup \Gamma^{out} \cup \Gamma \cup \{tick, stop\}$ 
    \item $\xi_{ac}: Q_{ac} \times \Sigma_{ac} \rightarrow Q_{ac}$
\end{itemize}

The (partial) transition function $\xi_{ac}$ is defined as follows:
\begin{enumerate}[1.]
    \setlength{\itemsep}{2pt}
    \setlength{\parsep}{0pt}
    \setlength{\parskip}{0pt}
    \item For any $\sigma \in \Sigma_{uo} \cup \Sigma_{o}^{out} \cup \Gamma^{in} \cup \Gamma^{out} \cup \Gamma \cup \{tick\}$, $\xi_{ac}(q_{ac}^{init}, \sigma) = q_{ac}^{init}$.
    \item For any $\sigma \in \Sigma_{o} - \Sigma_{o,a}$, $\xi_{ac}(q_{ac}^{init}, \sigma) = q_{uo}^{\sigma}$.
    \item For any $\sigma \in \Sigma_{o} - \Sigma_{o,a}$, $\xi_{ac}(q_{uo}^{\sigma}, \sigma^{in}) = q_{ac}^{init}$.
    \item For any $\sigma \in \Sigma_{s,a}$, $\xi_{ac}(q_{ac}^{init}, \sigma) = q_{0}$.
    \item For any $\sigma \in \Sigma_{o,a} - \Sigma_{s,a}$, $\xi_{ac}(q_{ac}^{init}, \sigma) = q_{o}^{\sigma}$.
    \item For any $\sigma \in \Sigma_{o,a} - \Sigma_{s,a}$, $\xi_{ac}(q_{o}^{\sigma}, \sigma^{in}) = q_{1}$. 
    \item For any $n \in [0:U]$, $\xi_{ac}(q_{n}, stop) = q_{ac}^{init}$.
    \item For any $n \in [0:U-1]$ and any $\sigma \in \Sigma_{s,a}$, $\xi_{ac}(q_{n}, \sigma^{\#}) = q_{n+1}$.
\end{enumerate}

\begin{figure}[htp]
\begin{center}
\includegraphics[height=3.2cm]{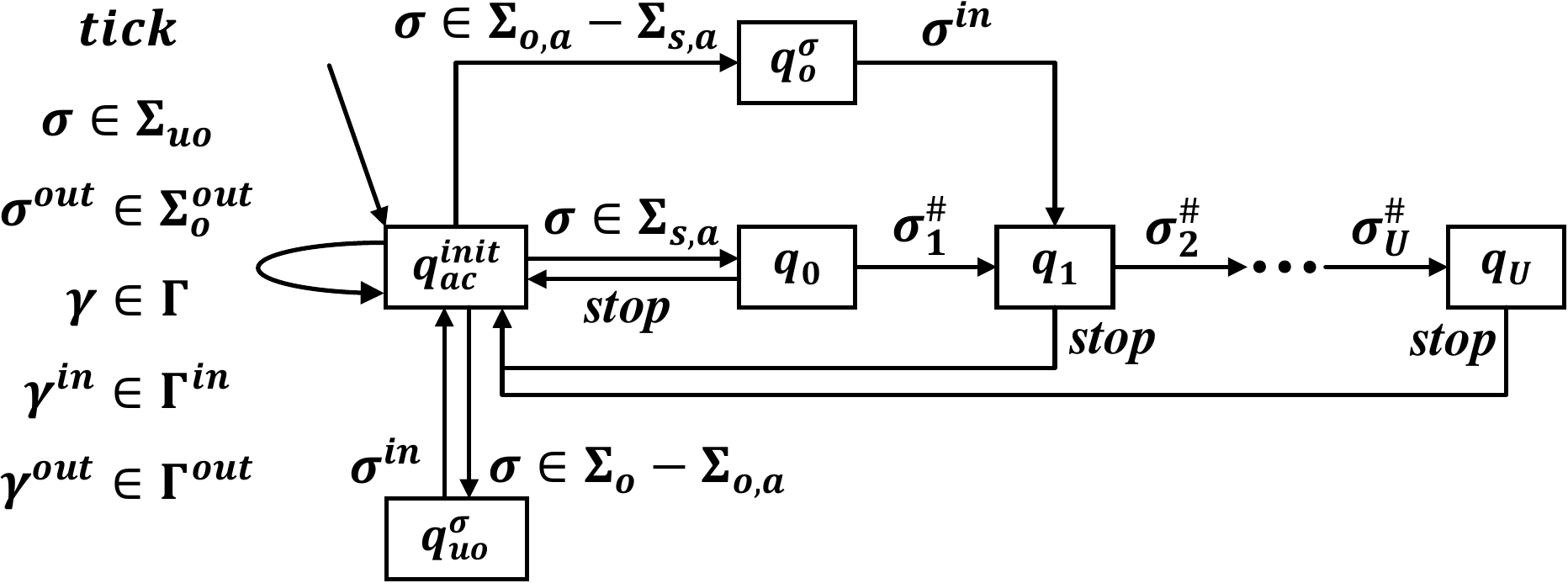}  
\caption{The (schematic) model for sensor attack constraints ($\sigma_{1}, \dots, \sigma_{U} \in \Sigma_{s,a}$)}
\label{fig:Sensor attack capability}
\end{center}                                
\end{figure}

The (schematic) model for sensor attack constraints $AC$ is shown in Fig. \ref{fig:Sensor attack capability},

Next, we shall present some explanations for the model $AC$. For the state set $Q_{ac}$, 
\begin{itemize}    
    \setlength{\itemsep}{3pt}
    \setlength{\parsep}{0pt}
    \setlength{\parskip}{0pt}
    \item $q_{o}^{\sigma}$ is a state denoting that $G$ executes an event $\sigma \in \Sigma_{o,a} - \Sigma_{s,a}$, which cannot be compromised, and the attacker has observed this event $\sigma$.
    \item $q_{uo}^{\sigma}$ is a state denoting that an event $\sigma \in \Sigma_{o} - \Sigma_{o,a}$ has been executed by $G$, which cannot be observed by the attacker.
    \item $q_{n} (n \in [0:U])$ is a state denoting that the attacker could either insert compromised events or stop inserting compromised events, i.e., end the attack. $n$ denotes the number of events that the attacker has already sent into the observation channel since it observes an event. Thus, at state $q_{n} (n \in [0:U])$, the number of events that the attacker could still send into the observation channel is $U-n$.
    \item $q_{ac}^{init}$ is a state denoting that either no attack has been conducted or the last attack has ended and since then the attacker has not observed any event in $\Sigma_{o}$.
\end{itemize}

In the event set $\Sigma_{ac}$, any $\sigma \in \Sigma$ denotes an event executed by plant $G$, any $\sigma^{in} \in (\Sigma_{o} - \Sigma_{s,a})^{in}$ denotes an event of sending an uncompromised event $\sigma \in \Sigma_{o} - \Sigma_{s,a}$ into the observation channel by the plant, and any $\sigma^{\#} \in \Sigma_{s,a}^{\#}$ denotes an event of sending a compromised event $\sigma \in \Sigma_{s,a}$ into the observation channel by the sensor attacker. The event $stop$ denotes the end of current round of attack, which are assumed to be controllable and observable, as they are initiated by the attacker.
Any element in $\Sigma_{o}^{out} \cup \Gamma^{in} \cup \Gamma^{out} \cup \Gamma$ denotes the event happening in the other five components: observation channel, networked supervisor, networked monitor, control channel, and plant with command execution and storage. All of the events in $\Sigma_{o}^{out} \cup \Gamma^{in} \cup \Gamma^{out} \cup \Gamma$ are unobservable to the sensor attacker.

For the (partial) transition function $\xi_{ac}$,
\begin{itemize}
    \setlength{\itemsep}{2pt}
    \setlength{\parsep}{0pt}
    \setlength{\parskip}{0pt}
    \item Case 1 says that, at state $q_{ac}^{init}$, 1) if any event $\sigma \in \Sigma_{uo} \cup \Sigma_{o}^{out} \cup \Gamma \cup \Gamma^{in} \cup \Gamma^{out}$ happens, then no attacks will be carried out since the attacker cannot observe $\sigma$; 2) $tick$ can always happen. All these events lead to a self-loop.
    \item Case 2 and 3 say that if an event $\sigma \in \Sigma_{o} - \Sigma_{o,a}$ is executed at $G$, then no attacks will be carried out because the attacker cannot observe $\sigma$. Since $\sigma$ is observable to the networked supervisor, we have the transitions %\footnote{The transition $\xi_{ac}(q_{ac}^{init}, \sigma) = q_{uo}^{\sigma}$ does not mean that the attack could observe the event $\sigma$. 
    %It is for technical convenience in the synchronous product of $AT$ and plant $G$ which will be introduced later in Section \ref{sec:Synthesis of Sensor attacks for Networked DES}
    %} 
     $\xi_{ac}(q_{ac}^{init}, \sigma) = q_{uo}^{\sigma}$, denoting that $\sigma$ has been executed, and $\xi_{ac}(q_{uo}^{\sigma}, \sigma^{in}) = q_{ac}^{init}$, denoting that $\sigma$ will be sent into the observation channel by the plant $G$. At state $q_{uo}^{\sigma}$, no other events would happen because when $\sigma$ is executed by the plant $G$, then it will be instantaneously sent into the observation channel.
    \item Case 4 says that after the attacker observes a compromised event $\sigma \in \Sigma_{s,a}$, it would transit to the state $q_{0}$, at which it could send a sequence of compromised events bounded by $U$ into the observation channel.
    \item Case 5 and 6 say that after the attacker observes $\sigma \in \Sigma_{o,a} - \Sigma_{s,a}$, it would transit to the state $q_{o}^{\sigma}$ and let $\sigma$ enter the observation channel since $\sigma$ is an uncompromised event. In addition, since the upper bound of events that the attacker can simultaneously send into the channel is $U$, the attacker could still send at most $(U-1)$ compromised events into the channel after observing $\sigma$\footnote{In this work, we shall count $\sigma$ in the events sent by the attacker. If we do not count $\sigma$, only minor modifications are needed, that is, replacing $\xi_{ac}(q_{o}^{\sigma}, \sigma^{in}) = q_{1}$ with $\xi_{ac}(q_{o}^{\sigma}, \sigma^{in}) = q_{0}$.}. Thus, we have the transitions $\xi_{ac}(q_{ac}^{init}, \sigma) = q_{o}^{\sigma}$ and $\xi_{ac}(q_{o}^{\sigma}, \sigma^{in}) = q_{1}$.
    \item Case 7 says that, at any state $q_{n} (n \in [0:U])$, the attacker could end the attack and transits back to the state $q_{ac}^{init}$.
    \item Case 8 says that at any state $q_{n} (n \in [0:U-1])$, the attacker could send any compromised event into the channel. Since the upper bound of events that the attacker can simultaneously send into the channel is $U$, the attacker cannot insert any compromised event at state $q_{U}$.
\end{itemize}
Based on the model of $AC$, the state size of $AC$ is $|Q_{ac}| = U + 2 + |\Sigma_{o} - \Sigma_{s,a}|$.

\textbf{Sensor attack:} A sensor attack over attack constraint $(\Sigma_{o,a}, \Sigma_{s,a})$ is modeled as a finite state automaton  
\[
A = (Q_{a}, \Sigma_{a}, \xi_{a}, q_{a}^{init}, Q_{a,m})
\] 
where $\Sigma_{a} = \Sigma_{ac} = \Sigma \cup (\Sigma_{o} - \Sigma_{s,a})^{in} \cup \Sigma_{s,a}^{\#} \cup \Sigma_{o}^{out} \cup \Gamma^{in} \cup \Gamma^{out} \cup \Gamma \cup \{tick, stop\}$, that satisfies the following constraints:
\vspace{-0.2cm}
\begin{itemize}
\setlength{\itemsep}{2pt}
\setlength{\parsep}{0pt}
\setlength{\parskip}{0pt}
    \item (SA-controllability) For any state $q \in Q_{a}$ and any event $\sigma \in \Sigma_{a,uc} = \Sigma_{a} - (\Sigma_{s,a}^{\#} \cup \{stop\})$, $\xi_{a}(q, \sigma)!$ 
    \item (SA-observability) For any state $q \in Q_{a}$ and any event $\sigma \in \Sigma_{a,uo} = \Sigma_{a} - (\Sigma_{o,a} \cup (\Sigma_{o,a} - \Sigma_{s,a})^{in} \cup \Sigma_{s,a}^{\#} \cup \{tick, stop\})$, if $\xi_{a}(q, \sigma)$!, then $\xi_{a}(q, \sigma) = q$.
\end{itemize}
SA-controllability states that the sensor attacker can only disable events in $\Sigma_{s,a}^{\#} \cup \{stop\}$. SA-observability states that the sensor attacker can only make a state change after observing an event 
%\footnote{In this work, we shall treat the events in $(\Sigma_{o,a} - \Sigma_{s,a})^{in} \cup \Sigma_{s,a}^{\#} \cup \{stop\}$ as being observable to the attack, although the opposite scenario can also be dealt with in a similar way.}
in $\Sigma_{o,a} \cup (\Sigma_{o,a} - \Sigma_{s,a})^{in} \cup \Sigma_{s,a}^{\#} \cup \{tick, stop\}$. In this work, by construction, all the controllable events for the attacker are also observable to the attacker. In the following text, we shall refer to 
\[
\begin{aligned}
\mathscr{C}_{ac} = (& \Sigma_{s,a}^{\#} \cup \{stop\}, \\ & \Sigma_{o,a} \cup (\Sigma_{o,a} - \Sigma_{s,a})^{in} \cup \Sigma_{s,a}^{\#} \cup \{tick, stop\})
\end{aligned}
\] 
as the attack-control constraint.

\subsection{Observation Channel}
\label{subsec:Observation Channel}
In Fig. \ref{fig:Networked supervisory control architecture under sensor attacks}, the observation channel is a module, representing a path for observations passing from the plant $G$ to the supervisor $S$, some of which may be intercepted and altered by the attacker. 
In the observation channel, delays may exist and practically, there always exists an upper bound of such delays, either due to a physical limit or a timeout mechanism set by the communication protocol. We assume that this upper bound is known and denoted as $\Delta_{o}$. At any discrete time $t \in \mathbb{N}$, the maximum\footnote{Under different event execution and transmission cases, for the same time instant, the number of messages transmitted in the channel might be different. Here, $N_{o}$ is the maximum among different cases.} number of events transmitted in the observation channel is defined as a mapping $N_{o}: \mathbb{N} \rightarrow \mathbb{N}$. To avoid the physically unrealistic possibility that plant $G$ will fire infinite events within a fixed unit time interval, i.e., one $tick$, we shall adopt a technical condition for the plant $G$, named \emph{activity-loop-freeness}, proposed in \cite{brandin1994supervisory}. For technical convenience, we shall also assume the largest number of fired events\footnote{Based on this assumption, the largest number of fired observable events within each tick at plant $G$ is no more than $N_{f}$.} within each tick at plant $G$ is $N_{f}$. Then we have the following theorem.

\emph{Theorem III.1:} Given $N_{f}$, $U$, and $\Delta_{o}$, for the observation channel, it holds that 
\[
\max\limits_{t \in \mathbb{N}}N_{o}(t) = N_{f}U(\Delta_{o}+1)
\]

\emph{Proof:} See Appendix A. \hfill $\blacksquare$

In the following text, we shall denote $C_{oc} = \max\limits_{t \in \mathbb{N}}N_{o}(t)$ as the observation channel capacity.
Based on Theorem III.1, the observation channel can be modeled as a finite state automaton 
\[
OC = (Q_{oc}, \Sigma_{oc}, \xi_{oc}, q_{oc}^{init})
\]
\begin{itemize}
    \setlength{\itemsep}{3pt}
    \setlength{\parsep}{0pt}
    \setlength{\parskip}{0pt}
    \item $Q_{oc} = \mathbb{N}^{\Sigma_{o} \times [0:\Delta_{o}]}(C_{oc})$ 
    \item $\Sigma_{oc} = (\Sigma_{o} - \Sigma_{s,a})^{in} \cup \Sigma_{s,a}^{\#} \cup \Sigma_{o}^{out} \cup \{tick\}$
    \item $\xi_{oc} \subseteq Q_{oc} \times \Sigma_{oc} \times Q_{oc}$
    \item $q_{oc}^{init} = \varnothing$
\end{itemize}

Before presenting the definition of the (partial) transition function $\xi_{oc}$, we define the following operations for two multisets $q, \hat{q} \in Q_{oc}$:
\begin{itemize}
    \setlength{\itemsep}{3pt}
    \setlength{\parsep}{0pt}
    \setlength{\parskip}{0pt}
    \item $T^{=0}(q) := \{(\sigma,t)^{m} \in q | t = 0\}$\footnote{$m$ is the multiplicity of the element $(\sigma,t)$ in $q$. For example, if a multiset $q = \{(a,1), (a,1), (b,1)\}$, then it can be written as $\{(a,1)^{2}, (b,1)\}$. The multiplicities of the elements $(a,1)$ and $(b,1)$ are 2 and 1, respectively, where the multiplicity is omitted if it is equal to 1.}
    \item $Tick(q) := \{(\sigma,t-1)^{m}|(\sigma,t)^{m} \in q\}$
    \item $\hat{q}$ is included in $q$, denoted as $\hat{q} \subseteq q$, if
        \begin{itemize}
            \item $(\forall (\sigma,t)^{m} \in \hat{q})(\exists n \in \mathbb{N}^{+}) \, (\sigma,t)^{n} \in q \wedge n \geq m$
        \end{itemize}
    \item $\uplus$ is the sum operation of multiplicities with the same base for multisets. $q \uplus \hat{q} = \{(\sigma,t)^{m+\hat{m}}|(\sigma,t)^{m} \in q \wedge (\sigma,t)^{\hat{m}} \in \hat{q}\} \cup 
    \{(\sigma,t)^{m} \in q | (\forall \hat{m} \in \mathbb{N}^{+})(\sigma,t)^{\hat{m}} \notin \hat{q}\} \cup
    \{(\sigma,t)^{\hat{m}} \in \hat{q} | (\forall m \in \mathbb{N}^{+})(\sigma,t)^{m} \notin q\}$
    \item If $\hat{q} \subseteq q$, then $q - \hat{q} := \{(\sigma,t)^{m-\hat{m}}|(\sigma,t)^{m} \in q \wedge (\sigma,t)^{\hat{m}} \in \hat{q}\} \uplus \{(\sigma,t)^{m} \in q | (\forall \hat{m} \in \mathbb{N}^{+})(\sigma,t)^{\hat{m}} \notin \hat{q}\}$
\end{itemize}

Then the transition relation $\xi_{oc}$ is defined as follows:
\begin{enumerate}[1.]
    \item For any $q \in Q_{oc}$ such that $T^{=0}(q) = \varnothing$, $(q, tick, Tick(q)) \\ \in \xi_{oc}$.
    \item For any $q \in Q_{oc}$ and any $\sigma \in \Sigma_{o} - \Sigma_{s,a}$, $(q, \sigma^{in}, q \uplus \{(\sigma, \Delta_{o})\}) \in \xi_{oc}$.
    \item For any $q \in Q_{oc}$ and any $\sigma \in \Sigma_{s,a}$, $(q, \sigma^{\#}, q \uplus \{(\sigma, \Delta_{o})\}) \in \xi_{oc}$.
    \item For any $\varnothing \neq q \in Q_{oc}$ and any $\sigma \in \Sigma_{o}$, if there exist $m \in \mathbb{N}^{+}$ and $n \in \mathbb{N}$ such that $(\sigma, n)^{m} \in q$, then $(q, \sigma^{out}, q - \{(\sigma, n)\}) \in \xi_{oc}$. 
    %\item Any other transition, except for those already defined in case 1, 2, 3, is not defined.
\end{enumerate}

We shall briefly explain the model $OC$. In the state set, except for the initial state $q_{oc}^{init} = \varnothing$, each state is a multiset of tuples. Each tuple consists of two components: 1) an observable event $\sigma$ transmitted in the channel\footnote{Technically, it is the message that encodes the event that is being transmitted in the channel.}; 2) the maximum number $t$ of time steps for which $\sigma$ could stay in the channel, that is, after $t$ tick events happen, $\sigma$ must be popped out from the observation channel, and at anytime during the gap of $t$ tick events, $\sigma$ could be popped out from the channel.
In the event set $\Sigma_{oc}$, any $\sigma^{out} \in \Sigma_{o}^{out}$ denotes an event of popping out an observable event from the observation channel.

For the transition relation $\xi_{oc}$,
\begin{itemize}
    \item Case 1 says that, for any state $q$ where all the second components of tuples are not equal to zero, tick event can happen, after which all the second components of tuples should minus one. For any state where there exists a tuple whose second component is equal to zero, tick event is not defined since now the observation channel must pop out some event.
    \item Case 2 says that, for any state $q$, if $\sigma^{in} \in (\Sigma_{o} - \Sigma_{s,a})^{in}$ happens, denoting that an event $\sigma$ is sent into the observation channel by the plant, then $OC$ will add this event into the channel attached with the maximum delay, denoted by $q \uplus \{(\sigma, \Delta_{o})\}$.
    \item Case 3 says that, for any state $q$, if $\sigma^{\#} \in \Sigma_{s,a}^{\#}$ happens, denoting that a compromised event $\sigma$ is sent into the observation channel by the attacker, then $OC$ will add this event into the channel attached with the maximum delay, denoted by $q \uplus \{(\sigma, \Delta_{o})\}$.
    \item Case 4 says that, for any state $q \neq \varnothing$, any event transmitted in the observation channel can be popped out, after which $OC$ will transit to the state $q - \{(\sigma, n)\}$, and this renders the observation channel non-FIFO and causes nondeterminism\footnote{Since the nondeterminism is not observable to the sensor attacker, nondeterminism can be subsumed by partial observation, and thus both TCT \cite{feng2006tct} and SuSyNA~\cite{SuSyNA} can be used for the synthesis.}. 
\end{itemize}
Based on the model of $OC$, the state size of $OC$ is $|Q_{oc}| = \frac{[|\Sigma_{o}|(\Delta_{o}+1)]^{C_{oc}+1}-1}{|\Sigma_{o}|(\Delta_{o}+1)-1}$.

\begin{figure}[htp]
\begin{center}
\includegraphics[height=1.4cm]{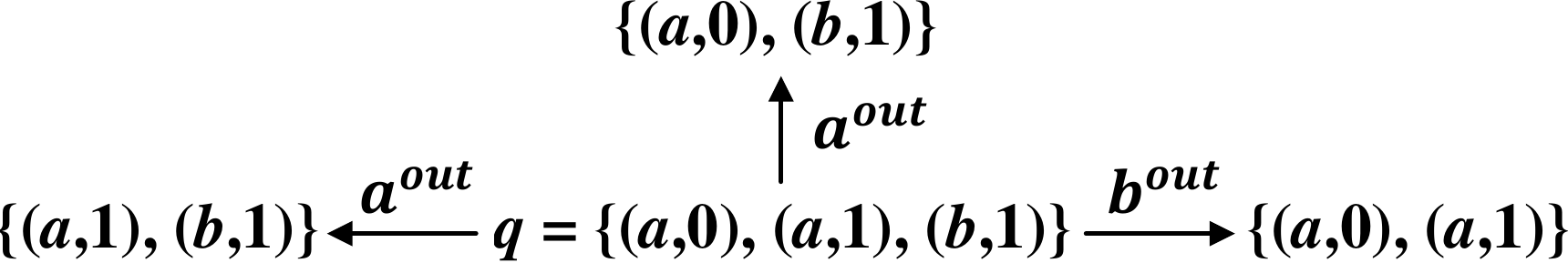}
\caption{Illustration of an example for case 4}
\label{fig:Illustration for case 4 of OC transition function}
\end{center}                                
\end{figure}

Next, we shall present a toy example in Fig. \ref{fig:Illustration for case 4 of OC transition function} to explain case 4 of $\xi_{oc}$. At state $q = \{(a,0),(a,1),(b,1)\}$, there are two events $a$ and one event $b$ transmitted in the observation channel. Thus, both $a$ and $b$ could be popped out. If $a$ is popped out, denoted by the transition labelled as $a^{out}$, then nondeterminism occurs since anyone of two events $a$ transmitted in the observation channel could be popped out, resulting in two possible states, $\{(a,0),(b,1)\}$ and $\{(a,1),(b,1)\}$. If $b$ is popped out, denoted by the transition labelled as $b^{out}$, then state $q$ will transit to state $\{(a,0),(a,1)\}$ without nondeterminism since there is only one event $b$ transmitted in the observation channel.

\subsection{Control Channel}
\label{subsec:Control Channel}
In Fig. \ref{fig:Networked supervisory control architecture under sensor attacks}, the control channel is a module, representing the directed message path from the networked supervisor to the plant with command execution and storage. In the control channel, delays may exist and the upper bound of the delay is $\Delta_{c}$. At any discrete time $t \in \mathbb{N}$, the maximum number of control commands transmitted in the control channel is defined as a mapping $N_{c}: \mathbb{N} \rightarrow \mathbb{N}$.

In this work, it is assumed that after observing any sequence consisting of events in $\Sigma_{o}^{out}$ and $tick$, the maximum number of control commands that can be sent by the networked supervisor is $V$. Then we have the following theorem.

\emph{Theorem III.2:} Given $N_{f}$, $U$, $V$, $\Delta_{o}$, and $\Delta_{c}$, for the control channel, it holds that 
\[
\max\limits_{t \in \mathbb{N}}N_{c}(t) = N_{f}UV(\Delta_{o} + \Delta_{c} + 1) + V(\Delta_{c} + 1)
\]

\emph{Proof:} See Appendix B. \hfill $\blacksquare$

In the following text, we shall denote $C_{cc} = \max\limits_{t \in \mathbb{N}}N_{c}(t)$ as the control channel capacity. Based on Theorem III.2, the control channel can be modeled as a finite state automaton 
\[
CC = (Q_{cc}, \Sigma_{cc}, \xi_{cc}, q_{cc}^{init})
\]
\begin{itemize}
    \setlength{\itemsep}{3pt}
    \setlength{\parsep}{0pt}
    \setlength{\parskip}{0pt}
    \item $Q_{cc} = \mathbb{N}^{\Gamma \times [0:\Delta_{c}]}(C_{cc})$ 
    \item $\Sigma_{cc} = \Gamma^{in} \cup \Gamma^{out} \cup \{tick\}$
    \item $\xi_{cc} \subseteq Q_{cc} \times \Sigma_{cc} \times Q_{cc}$
    \item $q_{cc}^{init} = \varnothing$
\end{itemize}

For any two multisets $q, \hat{q} \in Q_{cc}$, the operations $T^{=0}(q)$, $Tick(q)$, $\hat{q} \subseteq q$, $q \uplus \hat{q}$, and $q - \hat{q}$ follow the same definitions in Section \ref{subsec:Observation Channel}.
Then the transition relation $\xi_{cc}$ is defined as follows:
\begin{enumerate}[1.]
    \item For any $q \in Q_{cc}$ such that $T^{=0}(q) = \varnothing$, $(q, tick, Tick(q)) \\ \in \xi_{cc}$.
    \item For any $q \in Q_{cc}$ and any $\gamma \in \Gamma$, $(q, \gamma^{in}, q \uplus \{(\gamma, \Delta_{c})\}) \in \xi_{cc}$.
    \item For any $\varnothing \neq q \in Q_{cc}$ and any $\gamma \in \Gamma$, if there exist $m \in \mathbb{N}^{+}$ and $n \in \mathbb{N}$ such that $(\gamma, n)^{m} \in q$, then $(q, \gamma^{out}, q - \{(\gamma, n)\}) \in \xi_{cc}$.
\end{enumerate}

In the state set, except for the initial state $q_{cc}^{init} = \varnothing$, each state is a multiset of tuples. Each tuple consists of two components: 1) a control command $\gamma$ transmitted in the channel; 2) the maximum number $t$ of time steps for which $\gamma$ could stay in the channel. For the event set of the model $CC$, any $\gamma^{in}$ in $\Gamma^{in}$ denotes the event of sending a control command $\gamma$ into the control channel by the networked supervisor, and any $\gamma^{out} \in \Gamma^{out}$ denotes an event of popping out a control command $\gamma$ from the control channel. For the transition relation $\xi_{cc}$, it can be interpreted in a similar way as $\xi_{oc}$. By construction, the control channel is non-FIFO and the automaton model is nondeterministic.
Based on the model of $CC$, the state size of $CC$ is $|Q_{cc}| = \frac{[|\Gamma|(\Delta_{c}+1)]^{C_{cc}+1}-1}{|\Gamma|(\Delta_{c}+1)-1}$.

\subsection{Plant with Command Execution and Storage}
\label{Plant with Command Execution and Storage}

In this subsection, we shall construct the model of plant with command execution and storage. Before the formal definition, some assumptions used in this work are given as follows:

\begin{itemize}
\setlength{\itemsep}{3pt}
\setlength{\parsep}{0pt}
\setlength{\parskip}{0pt}
    \item There exists a gap\footnote{If $t_{e}^{\sigma} = 0$, then $\sigma$ is executed immediately at the time when $G$ starts to use the control command.} $t_{e}^{\sigma} \in \mathbb{N}$ between the time when $G$ starts to use one control command $\gamma \in \Gamma$ and  the time when an event $\sigma \in \gamma$ is fired at $G$, that is, only after $t_{e}^{\sigma}$ tick events happen, $\sigma$ could be fired\footnote{$t_{e}^{\sigma}$ is defined for any $\sigma \in \gamma \subseteq \Sigma_{c}$.}. Before $\sigma$ is executed, an uncontrollable event can also be executed, preempting the occurrence of $\sigma$.
    \item Uncontrollable events are always allowed to be fired at $G$ if there are uncontrollable events defined at the current state of $G$. 
    \item $G$ is able to store received control commands in the memory and the specific storage mechanism will be introduced later. Due to the limited memory, $G$ would not waste energy to store those very old control commands, based on which it is assumed in this work that there exists an upper bound of time steps, denoted as $\Delta_{s} \in \mathbb{N}$, for a control command that can be stored in the memory.
\end{itemize}
\begin{figure}[htp]
\begin{center}
\includegraphics[height=2.6cm]{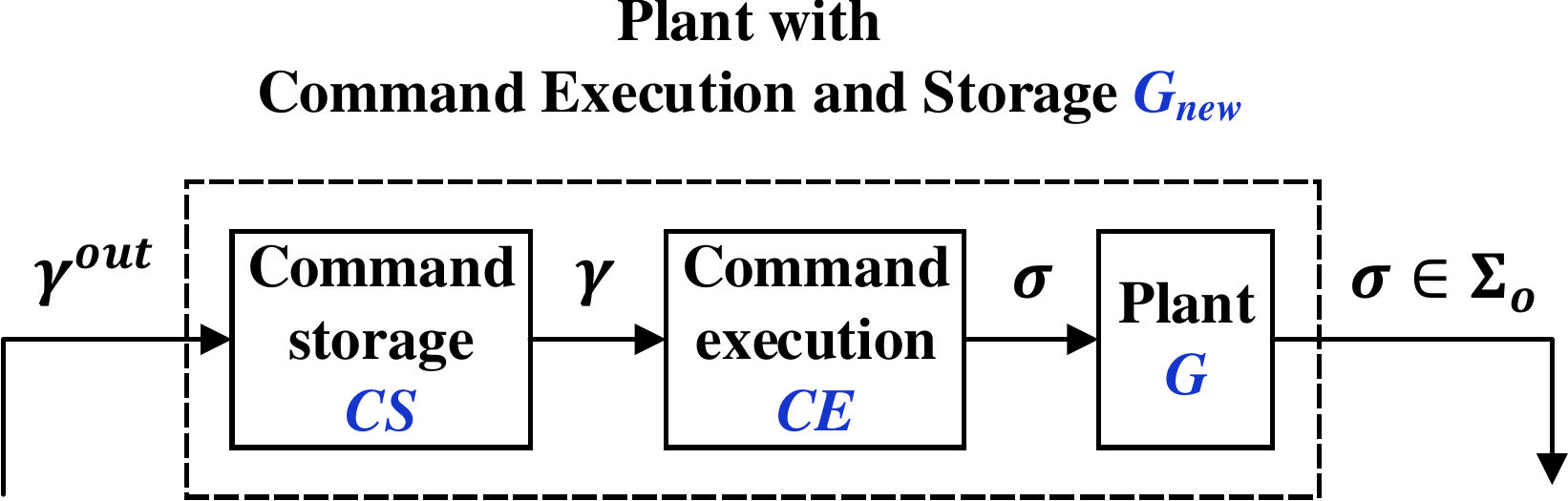}  
\caption{Internal structure of the plant with command execution and storage $G_{new}$}
\label{fig:Plant with command execution and storage}
\end{center}                                
\end{figure}

For the component $G_{new}$, which is the plant with command execution and storage shown in Fig. \ref{fig:Networked supervisory control architecture under sensor attacks}, its internal structure is illustrated in Fig. \ref{fig:Plant with command execution and storage}. There are three subcomponents: 

\begin{itemize}
\setlength{\itemsep}{3pt}
\setlength{\parsep}{0pt}
\setlength{\parskip}{0pt}
    \item Command storage automaton $CS$: It serves as the memory and stores a control command whenever it is received from the control channel. In this work, $CS$ is implemented as a FIFO queue and the technical details will be introduced later. The reasons for embedding this storage mechanism in the plant $G$ are explained as follows: On one hand, due to the non-FIFO property of the control channel, $G$ might receive in advance some control commands that should have been received later and cannot be used for the current state of $G$; on the other hand, $G$ might receive new control commands when it is using another control command. Thus, to deal with the above-mentioned two issues, in this work, we assume control commands can be stored.
    \item Command execution automaton $CE$: It explicitly describes the execution phase of the control command, that is, the procedure from using a control command till executing an event. The mechanism of the command execution is as follows: 
    \begin{enumerate}[1.]
    \setlength{\itemsep}{3pt}
    \setlength{\parsep}{0pt}
    \setlength{\parskip}{0pt}
        \item When it is not using any control command, it always tries to take out the earliest stored control command that can be used from the memory, i.e., $CS$. The control command taken out must contain some events defined at the current state of plant $G$.
        \item After it takes out one control command, it will not take out another one from $CS$ until some event is executed.
    \end{enumerate}  
    \item Plant $G$: It models the state change when an event is fired.
\end{itemize}

Then, the model of $G_{new}$ is constructed by computing the synchronous product of: 1) command storage automaton $CS$; 2) command execution automaton $CE$; 3) plant $G$, following by some pruning. Here, we shall firstly present some intuitive explanations for why we need the pruning procedure and the technical details will be introduced later. 
The pruning is needed because the synchronous product of $CS$, $CE$, and $G$ can not rule out the following two situations that are no consistent with the mechanism of $CE$ described above: 1) $CE$ fetches some control command, which is not usable for $G$, from $CS$; 2) $CE$ waits instead of fetching a control command that can be used from $CS$, leading to the phenomenon that some control command stored in $CS$, which could have been executed by $CE$, is erased from the memory when its storage time is up.

Next, we shall introduce how to formally construct the the model of plant with command execution and storage. In general, there are two construction steps:
\begin{enumerate}[1.]
    \item Construct the model of command storage $CS$, command execution $CE$, and plant $G$;
    \item Synthesize the plant with command execution and storage, named $G_{new}$, based on $CS$, $CE$, and $G$.
\end{enumerate}

\vspace{0.2cm}
\noindent \textbf{Step 1: Construct $CS$, $CE$, and $G$}

Intuitively speaking, $CS$ is a memory queue of control commands ordered by reception time. $CS$ always appends the received control command, which is popped out from the control channel, to the end of its memory queue and stores this control command for $\Delta_{s}$. The stored control commands in $CS$ are provided for $CE$ to use, that is, $CE$ could fetch control commands from $CS$ for event execution. 

At any discrete time $t \in \mathbb{N}$, the maximum number of control commands stored in the command storage module is defined as a mapping $N_{cs}: \mathbb{N} \rightarrow \mathbb{N}$. Then we have the following theorem.

\emph{Theorem III.3:} Given $N_{f}$, $U$, $V$, $\Delta_{o}$, $\Delta_{c}$, and $\Delta_{s}$, for the command storage module, it holds that 
\[
\max\limits_{t \in \mathbb{N}}N_{cs}(t) = N_{f}UV(\Delta_{o} + \Delta_{c} + \Delta_{s} + 1) + V(\Delta_{c} + \Delta_{s} + 1)
\]

\emph{Proof:} See Appendix C. \hfill $\blacksquare$

%{\color{red} early somewhere it can be explained that command storage is implemented as a FIFO queue}

In the following text, we shall denote $C_{cs} = \max\limits_{t \in \mathbb{N}}N_{cs}(t)$ as the command storage capacity. Based on Theorem III.3, the command storage can be modeled as a finite state automaton
\[
CS = (Q_{cs}, \Sigma_{cs}, \xi_{cs}, q_{cs}^{init})
\]
\begin{itemize}
\setlength{\itemsep}{3pt}
\setlength{\parsep}{0pt}
\setlength{\parskip}{0pt}
    \item $Q_{cs} = (\Gamma \times [0:\Delta_{s}])^{\leq C_{cs}}$  
    \item $\Sigma_{cs} = \Gamma^{out} \cup \Gamma \cup \{tick\}$
    \item $\xi_{cs}: Q_{cs} \times \Sigma_{cs} \rightarrow Q_{cs}$
    \item $q_{cs}^{init} = \varepsilon$
\end{itemize}

Before presenting the definition of the (partial) transition function $\xi_{cs}$, we define several operations as follows:
\begin{itemize}
\setlength{\itemsep}{3pt}
\setlength{\parsep}{0pt}
\setlength{\parskip}{0pt}
    \item $Tick: Q_{cs} \rightarrow Q_{cs}$ is a projection defined such that
    \begin{enumerate}[1.]
        \item $Tick(\varepsilon) = \varepsilon$,
        \item $(\forall (\gamma, t) \in (\Gamma \times [0:\Delta_{s}])) \\ Tick(\gamma, t)=
        \left\{
        \begin{array}{lcl}
        (\gamma, t-1)       &      & {t > 0,}\\
        \varepsilon  &      & {\rm otherwise.}
        \end{array} \right.$
        \item $(\forall s \in Q_{cs}, (\gamma, t) \in \Gamma \times [0:\Delta_{s}]) \, Tick(s(\gamma, t)) = Tick(s)Tick(\gamma, t)$. 
    \end{enumerate}
    \item For any $\varepsilon \neq q = (\gamma_{1}, t_{1})\dots(\gamma_{|q|}, t_{|q|}) =m_{1}\dots m_{|q|} \in Q_{cs}$,
    \begin{enumerate}[1.]
        \item $Com(q) := \{\gamma|(\exists t \in \mathbb{N})(\exists i \in [1:|q|])(\gamma,t) = m_{i}\}$.
        \item $Rem(q,\gamma) =\\
            \left\{
            \begin{array}{lcl}
            m_{1}\dots m_{i-1}m_{i+1}\dots m_{|q|}       &      & {(\exists i \in [1:|q|])(\forall 1 \leq }\\
                   &      & {j <i) \, \gamma_{j} \neq \gamma \wedge }\\
                   &      & {\gamma_{i} = \gamma}\\
            {\rm not \, defined}  &      & {\rm otherwise.}
            \end{array} \right.$
    \end{enumerate}
\end{itemize}
Intuitively speaking, for any tuple $(\gamma,t) \in \Gamma \times [0:\Delta_{s}]$, the second component $t$, if larger than zero, will minus one after $Tick$ operation. For any string $\varepsilon \neq q \in Q_{cs}$, $Com(q)$ extracts the first component of all the tuples in $q$ and $Rem(q,\gamma)$ removes the first tuple, whose first component is $\gamma$, in the state $q$.

Then the (partial) transition function $\xi_{cs}$ is defined as follows:
\begin{enumerate}[1.]
\setlength{\itemsep}{3pt}
\setlength{\parsep}{0pt}
\setlength{\parskip}{0pt}
    \item For any $q \in Q_{cs}$, $\xi_{cs}(q, tick) = Tick(q)$. 
    %where for $q = (\gamma_{1}, t_{1})\dots(\gamma_{|q|}, t_{|q|}) \in Q_{cs}$, $Tick(q) =\\
    %        \left\{
    %        \begin{array}{lcl}
    %        \varepsilon       &      & {q = \varepsilon}\\
    %        \varepsilon       &      & {(\forall 1 \leq i \leq |q|)t_{i} = 0}\\
    %        (\gamma_{1}, t_{1}-1)\dots(\gamma_{|q|}, t_{|q|}-1)       &      & {(\forall 1 \leq i \leq |q|)t_{i} \geq 1}\\
    %        (\gamma_{i}, t_{i}-1)\dots(\gamma_{|q|}, t_{|q|}-1)  &      & {(\exists 1 < i \leq |q|)t_{i-1} = }\\
    %          &      & {0 \wedge t_{i} \geq 1}\\
    %        {\rm not \, defined}  &      & {\rm otherwise.}
    %        \end{array} \right.
    %        $
    \item For any $q \in Q_{cs}$ and any $\gamma \in \Gamma$, $\xi_{cs}(q, \gamma^{out}) = q(\gamma, \Delta_{s}))$.
    \item For any $\varepsilon \neq q \in Q_{cs}$ and any $\gamma \in \Gamma$ such that $\gamma \in Com(q)$, $\xi_{cs}(q, \gamma) = Rem(q,\gamma)$, 
    %where
    %\begin{itemize}
    %\setlength{\itemsep}{3pt}
    %\setlength{\parsep}{0pt}
    %\setlength{\parskip}{0pt}
    %    \item $Com(q) = \{\gamma|(\exists t \in \mathbb{N})(\exists i \in [1:|q|])(\gamma,t) = m_{i}\}$. $Com(q)$ extracts all the control commands stored in the state $q$.
    %    \item $Rem(q,\gamma) =\\
    %        \left\{
    %        \begin{array}{lcl}
    %        m_{1}\dots m_{i-1}m_{i+1}\dots m_{|q|}       &      & {(\exists i \in [1:|q|])(\forall 1 \leq }\\
    %               &      & {j <i) \, \gamma_{j} \neq \gamma \wedge }\\
    %               &      & {\gamma_{i} = \gamma}\\
    %        {\rm not \, defined}  &      & {\rm otherwise.}
    %        \end{array} \right.$\\ 
    %        $Rem(q,\gamma)$ removes the first stored tuple containing $\gamma$ in the state $q$.
    %\end{itemize} 
\end{enumerate}

We shall briefly explain the model $CS$.
Thus, in the state set $Q_{cs}$, each state is a sequence of tuples. Each tuple contains the stored control command and the storage time, that is, the remaining time before this control command will be erased from the memory. In the event set $\Sigma_{cs}$, any $\gamma \in \Gamma$ denotes the event that $CS$ pops out the control command $\gamma$ for the command execution automaton $CE$, that is, $CE$ takes out the control command $\gamma$ to use from the memory.

For the (partial) transition function $\xi_{cs}$,
\begin{itemize}
\setlength{\itemsep}{3pt}
\setlength{\parsep}{0pt}
\setlength{\parskip}{0pt}
    \item Case 1 says that, after tick event happens, $CS$ will erase the stored control commands whose storage time is larger than $\Delta_{s}$ and the storage time of other stored control commands would minus one.
    \item Case 2 says that, once the event $\gamma^{out}$ happens, denoting that $CS$ receives a control command $\gamma$ popped out from the control channel, it will store the tuple $(\gamma, \Delta_{s})$, i.e., appending $(\gamma, \Delta_{s})$ to the end of $q$. 
    \item Case 3 says that, at any state $q \neq \varepsilon$, once the event $\gamma$ happens, then the first stored tuple $m_{i}$ containing $\gamma$ will be removed from the state $q$, denoted by $Rem(q,\gamma)$. This models the situation that $CE$ takes out the control command $\gamma$ from $CS$ to use.
\end{itemize}
Based on the model of $CS$, the state size of $CS$ is $|Q_{cs}| \leq \frac{[|\Gamma|(\Delta_{s}+1)]^{C_{cs}+1}-1}{|\Gamma|(\Delta_{s}+1)-1}$.

\begin{figure}[htp]
\begin{center}
\includegraphics[height=2.5cm]{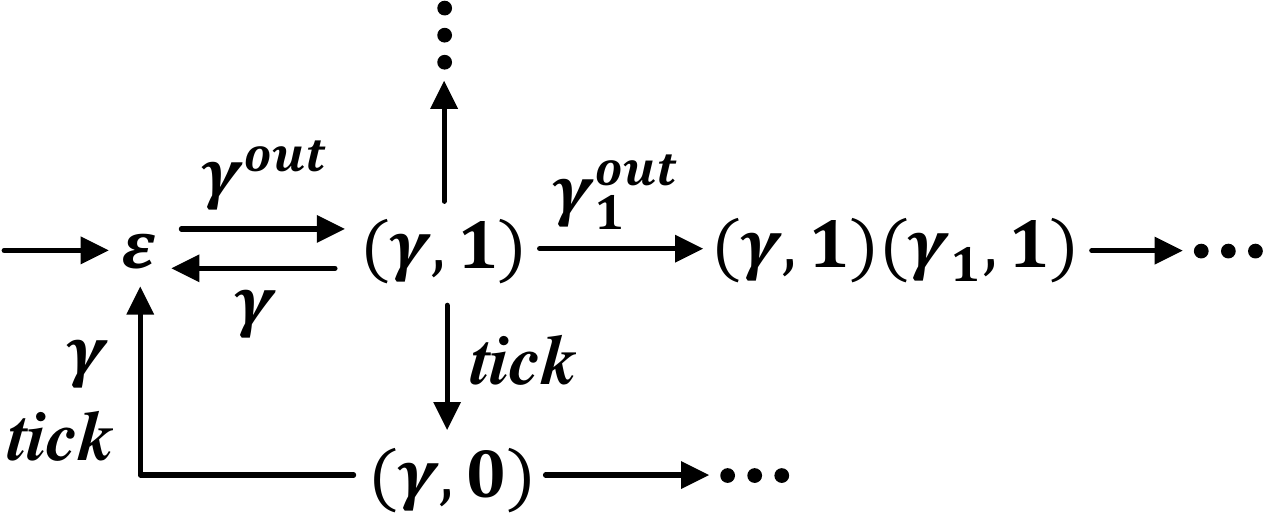}
\caption{Illustration of an example for $CS$}
\label{fig:Illustration for CS}
\end{center}                                
\end{figure}

Next, we shall present a toy example in Fig. \ref{fig:Illustration for CS} to explain the model $CS$. It is assumed $\Delta_{s} = 1$. At initial state $\varepsilon$, if a control command $\gamma$ is popped out from the control channel, denoted by the transition labelled as $\gamma^{out}$, then $CS$ will store it and transit to state $(\gamma,1)$. Afterwards, another control command $\gamma_{1}$ could also be popped out, resulting in the transition labelled as $\gamma_{1}^{out}$ to state $(\gamma,1)(\gamma_{1},1)$, where the new received control command $\gamma_{1}$ is appended to $\gamma$.
At state $(\gamma,1)$, if $CE$ fetches $\gamma$ from $CS$, denoted by the transition labelled as $\gamma$, then $CS$ will remove the earliest stored tuple containing $\gamma$ and transit back to initial state. At state $(\gamma,1)$, it is also possible that $CE$ takes no action, i.e., no control command is fetched, resulting in that only one $tick$ happens and $CS$ transits to state $(\gamma,0)$. Similarly, at state $(\gamma,0)$, any transition labelled as $\gamma$ or $tick$ will result in a transition to the initial state. 

Then we construct the command execution automaton $CE$. Before giving the formal definition, we shall define several operations as follows, for any $\gamma \in \Gamma$:
\begin{itemize}
\setlength{\itemsep}{3pt}
\setlength{\parsep}{0pt}
\setlength{\parskip}{0pt}
    \item $\gamma^{num} := \{(\sigma, t_{e}^{\sigma})|\sigma \in \gamma\}$.
    \item $Tick(\gamma^{num}, n) := \\
            \left\{
            \begin{array}{lcl}
            \{(\sigma,t-n)|(\sigma,t) \in \gamma^{num}\}       &      & {\rm if} \, {n \in [0:\mathop {\max}\limits_{\sigma \in \gamma}\{t_{e}^{\sigma}\}],}\\
            {\rm not \, defined}  &      & {\rm otherwise.}
            \end{array} \right.$\\
        %If $n = 1$, then we shall also write $Tick(\gamma^{num}, 1)$ as $Tick(\gamma^{num})$ for simplicity.
\end{itemize}

Then the command execution is modeled as a finite state automaton %{\color{red} I skip this part..}
\[
CE = (Q_{ce}, \Sigma_{ce}, \xi_{ce}, q_{ce}^{init})
\]
\begin{itemize}
\setlength{\itemsep}{3pt}
\setlength{\parsep}{0pt}
\setlength{\parskip}{0pt}
    \item $Q_{ce} = \{\varnothing\} \cup \bigcup\limits_{\gamma \in \Gamma} \bigcup\limits_{n \in [0:\mathop {\max}\limits_{\sigma \in \gamma}\{t_{e}^{\sigma}\}]} Tick(\gamma^{num}, n)$
    \item $\Sigma_{ce} = \Gamma \cup \Sigma \cup \{tick\}$
    \item $\xi_{ce}: Q_{ce} \times \Sigma_{ce} \rightarrow Q_{ce}$
    \item $q_{ce}^{init} = \varnothing$
\end{itemize}

Before presenting the definition of the (partial) transition function $\xi_{ce}$, we shall define several operations as follows, for any state $q \in Q_{ce}$:
\begin{itemize}
\setlength{\itemsep}{3pt}
\setlength{\parsep}{0pt}
\setlength{\parskip}{0pt}
    \item $T^{>0}(q) = \{\sigma|(\sigma,t) \in q \wedge t > 0\}$
    \item $T^{=0}(q) = \{\sigma|(\sigma,t) \in q \wedge t = 0\}$
    \item $Tick(q) = \{(\sigma,t-1)|(\sigma,t) \in q\}$
\end{itemize}

Then the (partial) transition function $\xi_{ce}$ is defined as follows:
\begin{enumerate}[1.]
\setlength{\itemsep}{3pt}
\setlength{\parsep}{0pt}
\setlength{\parskip}{0pt}
    \item For any $q \in Q_{ce}$ such that $q =\varnothing$ or $T^{>0}(q) \neq \varnothing$, $\xi_{ce}(q, tick) = Tick(q)$.
    \item For any $\gamma \in \Gamma$, $\xi_{ce}(q_{ce}^{init}, \gamma) = \gamma^{num}$.
    \item For any $\varnothing \neq q \in Q_{ce}$ such that $T^{=0}(q) \neq \varnothing$ and any $\sigma \in T^{=0}(q)$, $\xi_{ce}(q, \sigma) = q_{ce}^{init}$.
    \item For any $q \in Q_{ce}$ and any $\sigma \in \Sigma_{uc}$, $\xi_{ce}(q, \sigma) = q_{ce}^{init}$.
\end{enumerate}

We shall briefly explain the model $CE$. In the state set $Q_{ce}$, except for the initial state $q_{ce}^{init} = \varnothing$, each state is a set of tuples. In each tuple, the first component corresponds to an event, belonging to the control command being used by $CE$, and the second component, if nonnegative, corresponds to the remaining time before this event could be fired by $CE$. 

For the (partial) transition function $\xi_{ce}$,
\begin{itemize}
\setlength{\itemsep}{3pt}
\setlength{\parsep}{0pt}
\setlength{\parskip}{0pt}
    \item Case 1 says that at the initial state or any state $q$ which contains a tuple whose second component is larger than zero, $tick$ is defined. After tick event happens, all the second components of tuples in $q$ should minus one. For any state in $Q_{ce}$, if all the second components of tuples in this state are not larger than zero, then based on the definition of $Q_{ce}$, we know that the second components of some tuples in this state must be zero and the second components of the rest tuples in this state must be smaller than zero, in this case, $tick$ is not defined because $CE$ must fire an event now.
    \item Case 2 says that, at initial state, $CE$ could take out one stored control command $\gamma \in \Gamma$ from the memory to use, after which $CE$ would record the enabled events in $\gamma$ and the remaining time before each enabled event could be executed, denoted by $\gamma^{num}$.
    \item Case 3 says that, at any state not equal to $\varnothing$, $CE$ could execute an event $\sigma$ and transits to $q_{ce}^{init}$ only when 1) $\sigma$ belongs to the control command being used by $CE$; and 2) the remaining time before $\sigma$ could be executed becomes zero.
    \item Case 4 says that, at any state, uncontrollable events can always be fired, after which $CE$ would transit to the initial state $q_{ce}^{init}$.
\end{itemize}
Based on the model of $CE$, the state size of $CE$ is $|Q_{ce}| \leq |\Gamma|(1 + \mathop {\max}\limits_{\sigma \in \Sigma_{c}}\{t_{e}^{\sigma}\})$.

Finally, the plant $G$ is modeled as a finite state automaton $G = (Q, \Sigma, \xi, q_{0}, Q_{mark})$. We use $Q_{d}$ to denote the set of damage states, any state of which is a goal state that the sensor attacker targets to induce the plant $G$ to reach.

\vspace{0.2cm}
\noindent \textbf{Step 2: Synthesize $G_{new}$}

\vspace{0.1cm}

Firstly, based on $CS$, $CE$, and $G$, we construct 
\[G_{temp} = CS||CE||G = (Q_{temp}, \Sigma_{temp}, \xi_{temp}, q_{temp}^{init})\]
\begin{itemize}
\setlength{\itemsep}{3pt}
\setlength{\parsep}{0pt}
\setlength{\parskip}{0pt}
    \item $Q_{temp} = Q_{cs} \times Q_{ce} \times Q$
    \item $\Sigma_{temp} = \Sigma_{cs} \cup \Sigma_{ce} \cup \Sigma = \Gamma^{out} \cup \Gamma \cup \Sigma \cup \{tick\}$
    \item $\xi_{temp}: Q_{temp} \times \Sigma_{temp} \rightarrow Q_{temp}$
    \item $q_{temp}^{init} = (q_{cs}^{init}, q_{ce}^{init}, q_{0})$
\end{itemize}

Next, we shall generate the model of plant with command execution and storage, $G_{new}$, by implementing a pruning procedure on $G_{temp}$, which is defined as follows: %{\color{red} instead of removing states, it can be thought of removing transitions}

Let $G_{new} := (Q_{new}, \Sigma_{new}, \xi_{new}, q_{new}^{init})$
\begin{enumerate}[1.]
\setlength{\itemsep}{3pt}
\setlength{\parsep}{0pt}
\setlength{\parskip}{3pt}
    \item $Q_{new} := Q_{temp} - Q_{del}$, $Q_{del} = \{q_{temp} = (s,\gamma,q) \in Q_{temp} = Q_{cs} \times Q_{ce} \times Q|En_{G}(q) \cap \gamma^{denum} = \varnothing\}$, where for any $\gamma \in Q_{ce}$, $\gamma^{denum} = \{\sigma|(\exists t \in \mathbb{Z})(\sigma,t) \in \gamma\}$
    \item $\Sigma_{new} = \Sigma_{temp}$
    \item $(\forall q_{1}, q_{2} \in Q_{new})(\forall \sigma \in \Sigma_{temp} - \{tick\}) \, \xi_{temp}(q_{1}, \sigma) = q_{2} \Leftrightarrow \xi_{new}(q_{1}, \sigma) = q_{2}$
    \item $(\forall q_{1} = (s,\gamma,q), q_{2} \in Q_{new}) \, \xi_{temp}(q_{1}, tick) = q_{2} \wedge \neg C \Leftrightarrow \xi_{new}(q_{1}, tick) = q_{2}$, where $C := \gamma = q_{ce}^{init} = \varnothing \wedge (\exists \hat{\gamma} \in Com(s))En_{G}(q) \cap \hat{\gamma} \neq \varnothing$
    \item $q_{new}^{init} = q_{temp}^{init}$
\end{enumerate}

In the above pruning procedure, in Step 1, we need to delete those states in $Q_{del}$, any state of which satisfies the condition that, $CE$ is using a control command $\gamma^{denum}$ and $\gamma^{denum}$ does not include events that are defined at state $q$ of plant $G$, denoted by $En_{G}(q) \cap \gamma^{denum} = \varnothing$.
In Step 4, we need to delete the transitions labelled as $tick$ starting from state $q_{1} = (s, \gamma, q)$ that satisfies the following condition: 1) $CE$ is not using any control command, denoted by $\gamma = q_{ce}^{init}$; 2) there exists some control command $\hat{\gamma}$ stored in $CS$, denoted by $\hat{\gamma} \in Com(s)$; 3) $\hat{\gamma}$ has shared events with the enabled events defined at state $q$ of plant $G$, denoted by $En_{G}(q) \cap \hat{\gamma} \neq \varnothing$. In this case, based on the mechanism of the command execution module, $CE$ will fetch some control command that can be used from $CS$ instead of waiting, i.e., time is preempted. Thus, the transitions labelled as $tick$ satisfying the above condition would be deleted. 

\subsection{Networked Supervisor}
\label{subsec:Networked Supervisor}

The networked supervisor over control constraint $(\Gamma^{in}, \Gamma^{in} \cup \Sigma_{o}^{out} \cup \{tick\})$ is a finite state automaton 
\[
NS = (Q_{ns}, \Sigma_{ns}, \xi_{ns}, q_{ns}^{init})
\]
\begin{itemize}
    \item $\Sigma_{ns} = \Sigma \cup (\Sigma_{o} - \Sigma_{s,a})^{in} \cup \Sigma_{s,a}^{\#} \cup \Sigma_{o}^{out} \cup \Gamma^{in} \cup \Gamma^{out} \cup \Gamma \cup \{tick, stop\}$   
\end{itemize}

The following constraints should be satisfied: 
\begin{itemize}
\setlength{\itemsep}{3pt}
\setlength{\parsep}{0pt}
\setlength{\parskip}{0pt}
    \item (Network controllability) For any state $q \in Q_{ns}$ and any $\sigma \in \Sigma_{ns} - \Gamma^{in}$, $\xi_{ns}(q, \sigma)!$
    \item (Network observability) For any state $q \in Q_{ns}$ and any $\sigma \in \Sigma_{ns} - (\Gamma^{in} \cup \Sigma_{o}^{out} \cup \{tick\})$, if $\xi_{ns}(q, \sigma)!$, then $\xi_{ns}(q,\sigma) = q$.
\end{itemize}

%{\color{red} in this work, we shall assume the networked supervisor is given}
\subsection{Networked Monitor}
\label{subsec: Monitor}
In this work, we assume the networked supervisor $NS$ is augmented with a monitoring mechanism which monitors the execution of the closed-loop system and serves to detect the existence of a sensor attack. This monitoring mechanism is modeled as the networked monitor $M$ shown in Fig. \ref{fig:Networked supervisory control architecture under sensor attacks}. The networked monitor online records its observation sequence in $(\Sigma_{o}^{out} \cup \Gamma^{in} \cup \{tick\})^{*}$ 
%{\color{red} does the networked monitor also see the control commands it sent into the channel in Fig. 1? For networked systems, it may be critical to see the control commands..} 
of the closed-loop system (possibly under sensor attack) and determine the existence of a sensor attack based on its observation.
Intuitively, the principle of the networked monitor is that it always compares the observed sequence with the ones that should have been observed under the absence of a sensor attack. Once an information inconsistency happens, the networked monitor would identify it. 

Next, we shall construct the model of $M$. Firstly, since $OC$ given in Section \ref{subsec:Observation Channel} is the model of the observation channel under sensor attack and the detection mechanism of the networked monitor is based on a comparison with the observation sequence without sensor attacks, to construct the model of the networked monitor, we shall carry out some event relabelling on $OC$. 
Then, the model of the networked monitor $M$ is generated by computing the synchronous product, performing projection, and adding a self-loop labelled by $tick$, as shown in the following.

%For $G_{new}$, by replacing any transition labelled as $\sigma \in \Sigma_{o}$ with $\sigma^{in}$, the transformed plant with command execution and storage is generated and denoted as
%\[
%G_{new}^{T} = (Q_{new}, \Sigma_{new}^{T}, \xi_{new}^{T}, q_{new}^{init})
%\]
%\begin{itemize}
%\setlength{\itemsep}{3pt}
%\setlength{\parsep}{0pt}
%\setlength{\parskip}{0pt}
%    \item $\Sigma_{new}^{T} =\Gamma^{out} \cup \Gamma \cup (\Sigma - \Sigma_{o}) \cup \Sigma_{o}^{in} \cup \{tick\}$
%    \item $\xi_{new}^{T}: Q_{new} \times \Sigma_{new}^{T} \rightarrow Q_{new}$
%    \item $(\forall q, q^{'} \in Q_{new})(\forall \sigma \in \Sigma_{o}) \, \xi_{new}(q, \sigma) = q^{'} \Leftrightarrow \xi_{new}^{T}(q, \sigma^{in}) = q^{'}$
%    \item $(\forall q, q^{'} \in Q_{new})(\forall \sigma \in \Sigma_{new} - \Sigma_{o}) \, \xi_{new}(q, \sigma) = q^{'} \Leftrightarrow \xi_{new}^{T}(q, \sigma) = q^{'}$
%\end{itemize}

For $OC$, by replacing any transition labelled as $\sigma^{\#} \in \Sigma_{s,a}^{\#}$ with $\sigma$, and any transition labelled as $\sigma^{in} \in (\Sigma_{o} - \Sigma_{s,a})^{in}$ with $\sigma$, the transformed model of observation channel is generated and denoted as
\[
OC^{T} = (Q_{oc}, \Sigma_{oc}^{T}, \xi_{oc}^{T}, q_{oc}^{init})
\]
\begin{itemize}
\setlength{\itemsep}{3pt}
\setlength{\parsep}{0pt}
\setlength{\parskip}{0pt}
    \item $\Sigma_{oc}^{T} = \Sigma_{o} \cup \Sigma_{o}^{out} \cup \{tick\}$
    \item $\xi_{oc}^{T} \subseteq Q_{oc} \times \Sigma_{oc}^{T} \times Q_{oc}$
    \begin{enumerate}[1.]
        \item $(\forall q_{1}, q_{2} \in Q_{oc})(\forall \sigma \in \Sigma_{s,a}) \, (q_{1}, \sigma^{\#}, q_{2}) \in \xi_{oc} \Leftrightarrow (q_{1}, \sigma, q_{2}) \in \xi_{oc}^{T}$
        \item $(\forall q_{1}, q_{2} \in Q_{oc})(\forall \sigma \in \Sigma_{o} - \Sigma_{s,a}) \, (q_{1}, \sigma^{in}, q_{2}) \in \xi_{oc} \Leftrightarrow (q_{1}, \sigma, q_{2}) \in \xi_{oc}^{T}$
        \item $(\forall q_{1}, q_{2} \in Q_{oc})(\forall \sigma \in \Sigma_{oc} - \Sigma_{s,a}^{\#} - (\Sigma_{o} - \Sigma_{s,a})^{in}) \, (q_{1}, \sigma, q_{2}) \in \xi_{oc} \Leftrightarrow (q_{1}, \sigma, q_{2}) \in \xi_{oc}^{T}$
    \end{enumerate}
\end{itemize}

Based on the above-mentioned monitoring mechanism, the networked monitor would detect the existence of a sensor attack when it observes some string $s \notin P_{\Sigma_{o}^{out} \cup \Gamma^{in} \cup \{tick\}}(L(NS||G_{new}||OC^{T}||CC))$. Thus, based on $NS$, $G_{new}$, $OC^{T}$, and $CC$, the networked monitor is generated and denoted as   
\[
\begin{aligned}
M & = P_{\Sigma_{o}^{out} \cup \Gamma^{in} \cup \{tick\}}(NS||G_{new}||OC^{T}||CC) \\ & = (Q_{m}, \Sigma_{m}, \xi_{m}, q_{m}^{init}) 
\end{aligned}
\]
\begin{itemize}
\setlength{\itemsep}{3pt}
\setlength{\parsep}{0pt}
\setlength{\parskip}{0pt}
    \item $Q_{m} = 2^{Q_{ns} \times Q_{new} \times Q_{oc} \times Q_{cc}}$
    \item $\Sigma_{m} = \Sigma_{ns} \cup \Sigma_{new} \cup \Sigma_{oc}^{T} \cup \Sigma_{cc}$
    \item $\xi_{m}: Q_{m} \times \Sigma_{m} \rightarrow Q_{m}$ 
    \item $q_{m}^{init} = UR_{NS||G_{new}||OC^{T}||CC, \Sigma_{m} - (\Sigma_{o}^{out} \cup \Gamma^{in} \cup \{tick\})}(q_{ns}^{init}\\, q_{new}^{init}, q_{oc}^{init}, q_{cc}^{init})$
\end{itemize}
Based on the construction of $M$, once the monitor observes some event that should not have occurred, it will transit to state $\varnothing$, meaning that the sensor attack is detected. Thus, the covert sensor attack needs to avoid such transition in $M$.
Since tick event can still happen even if the attack has been detected, at state $\varnothing$ of $M$, we need to add the self-loop labelled by $tick$. Based on the model of $M$, the state size of $M$ is $|Q_{m}| \leq 2^{|Q_{cs}| \times |Q_{ce}| \times |Q| \times |Q_{oc}| \times |Q_{ns}| \times |Q_{cc}|}$.

%%%%%%%%%%%%%%%%%%%%%%%%%%%%%%%%%%%%%%%%%%%%%%%%%%%%%%%%%%%%%%%%%%%%%%%%%%%%%%%%
\vspace{0.3cm}

\section{Synthesis of Covert Sensor attacks for Networked DES}
\label{sec:Synthesis of Covert Sensor attacks for Networked DES}

In this section, 
%firstly, based on the component models presented in Section \ref{sec:Component Models for Networked DES under sensor attacks}, we shall formalize the closed-loop behavior of the networked DES under sensor attack. Based on the closed-loop behavior, we shall introduce several definitions, including covertness, damage-nonblocking property, and damage-reachable property. Then, 
we shall solve the synthesis problem of covert sensor attacks for networked DES by modeling it as the Ramadge-Wonham supervisory control problem. 
%Finally, the computational complexity analysis is presented.

%{\color{red} it is non-determinsitic transition structure, but since nondeterminism can be subsumed by partial observation, can use TCT or SuSYNA or Supremica, if do not consider forcing in the attack, currently, seems no need to consider forcing. great!}

\subsection{Solution Methodology}
\label{subsec: Transformation Methodology}
In Fig. \ref{fig:Networked supervisory control architecture under sensor attacks}, given the plant with command execution and storage $G_{new}$, the sensor attack constraints $AC$, the observation channel $OC$, the networked supervisor $NS$, the control channel $CC$, the networked monitor $M$, and the sensor attack $A$, the closed-loop behavior of the networked DES under sensor attack is the synchronous product 
\[
\begin{aligned}
\mathcal{B} & = G_{new}||AC||OC||NS||CC||M||A \\ & = (Q_{b}, \Sigma_{b}, \xi_{b}, q_{b}^{init}, Q_{b,m})
\end{aligned}
\]
\begin{itemize}
\setlength{\itemsep}{3pt}
\setlength{\parsep}{0pt}
\setlength{\parskip}{0pt}
    \item $Q_{b} = Q_{new} \times Q_{ac} \times Q_{oc} \times Q_{ns} \times Q_{cc} \times Q_{m} \times Q_{a}$
    \item $\Sigma_{b} = \Sigma_{new} \cup \Sigma_{ac} \cup \Sigma_{oc} \cup \Sigma_{ns} \cup \Sigma_{cc} \cup \Sigma_{m} \cup \Sigma_{a}$
    \item $\xi_{b} \subseteq Q_{b} \times \Sigma_{b} \times Q_{b}$
    \item $q_{b}^{init} = (q_{new}^{init}, q_{ac}^{init}, q_{oc}^{init}, q_{ns}^{init}, q_{cc}^{init}, q_{m}^{init}, q_{a}^{init})$
    \item $Q_{b,m} = \{((s, \gamma, q), q_{ac}, q_{oc}, q_{ns}, q_{cc}, q_{m}, q_{a}) \in Q_{b}|\,q \in Q_{d}\}$ 
\end{itemize}
In this work, we shall assume that the goal of a sensor attack is achieved when $G$ reaches the damage state in $Q_{d}$. Thus, from the point view of the sensor attack, its target state set in $\mathcal{B}$ is $Q_{b,m}$.

\emph{Definition IV.1. (Covertness):} Given any $G_{new}$, $AC$, $OC$, $NS$, $CC$, and $M$, the sensor attack $A$ is said to be covert w.r.t. the attack constraint $(\Sigma_{o,a}, \Sigma_{s,a})$ if any state in 
\[
\begin{aligned}
& Q_{bad} = \{((s,\gamma,q), q_{ac}, q_{oc}, q_{ns}, q_{cc}, q_{m}, q_{a}) \in Q_{b}|\, q \notin Q_{d} \\ & \wedge q_{m} = \varnothing\}
\end{aligned}
\]
is not reachable in $\mathcal{B}$.

\emph{Definition IV.2. (Damage-nonblocking):} Given any $G_{new}$, $AC$, $OC$, $NS$, $CC$, and $M$, the sensor attack $A$ is said to be damage-nonblocking (a strong attack) w.r.t. the attack constraint $(\Sigma_{o,a}, \Sigma_{s,a})$ if $\mathcal{B}$ is nonblocking, that is, every reachable state is coreachable, i.e., every reachable state can reach a marker state. 

\emph{Definition IV.3. (Damage-reachable):} Given any $G_{new}$, $AC$, $OC$, $NS$, $CC$, and $M$, the sensor attack $A$ is said to be damage-reachable (a weak attack) w.r.t. the attack constraint $(\Sigma_{o,a}, \Sigma_{s,a})$ if some state of $Q_{b,m}$ is reachable in $\mathcal{B}$, that is, $L_{m}(\mathcal{B}) \neq \varnothing$.

\vspace{0.1cm}

%\emph{Definition IV.4. (Supremal covert damage-nonblocking sensor attack):} Given any $G_{new}$, $AC$, $OC$, $NS$, $CC$, and $M$, a covert damage-nonblocking sensor attack $A$ is said to be the supremal covert damage-nonblocking sensor attack w.r.t. the attack constraint $(\Sigma_{o,a}, \Sigma_{s,a})$ if for any covert damage-nonblocking sensor attack $A^{'}$, it always holds that $L_{m}(G_{new}||AC||OC||NS||CC||M||A^{'}) \subseteq L_{m}(G_{new}||AC||OC||NS||CC||M||A)$.

%\vspace{0.1cm}

%\emph{Definition IV.5. (Supremal covert damage-reachable sensor attack):} Given any $G_{new}$, $AC$, $OC$, $NS$, $CC$, and $M$, a covert damage-reachable sensor attack $A$ is said to be the supremal covert damage-reachable sensor attack w.r.t. the attack constraint $(\Sigma_{o,a}, \Sigma_{s,a})$ if for any covert damage-reachable sensor attack $A^{'}$, it always holds that $L_{m}(G_{new}||AC||OC||NS||CC||M||A^{'}) \subseteq L_{m}(G_{new}||AC|| \\ OC||NS||CC||M||A)$ and $L(G_{new}||AC||OC||NS||CC||M|| \\ A^{'}) \subseteq L(G_{new}||AC||OC||NS||CC||M||A)$.

Next, we shall introduce the approach of modeling the synthesis problem of covert sensor attacks for networked DES as the Ramadge-Wonham supervisory control problem.
Since the networked DES under sensor attack is $\mathcal{B} = G_{new}||AC||OC||NS||CC||M||A$, we can view
\[
P = G_{new}||AC||OC||NS||CC||M = (Q_{p}, \Sigma_{p}, \xi_{p}, q_{p}^{init}, Q_{p,m})
\]
as the new plant and $A$ as the new supervisor to be synthesized, based on which we have the following results.

\emph{Theorem IV.1:} Given any $G_{new}$, $AC$, $OC$, $NS$, $CC$, and $M$, there exists a covert damage-nonblocking sensor attack $A$ w.r.t. the attack constraint $(\Sigma_{o,a}, \Sigma_{s,a})$ if and only if there exists a supervisor $S^{'}$ over attack-control constraint $\mathscr{C}_{ac} = (\Sigma_{s,a}^{\#} \cup \{stop\}, \Sigma_{o,a} \cup (\Sigma_{o,a} - \Sigma_{s,a})^{in} \cup \Sigma_{s,a}^{\#} \cup \{tick, stop\})$ such that 
\begin{itemize}
\setlength{\itemsep}{3pt}
\setlength{\parsep}{0pt}
\setlength{\parskip}{0pt}
    \item Any state in $Q_{bad}$ is not reachable in $P||S^{'}$.
    \item $P||S^{'}$ is nonblocking w.r.t. $Q_{b,m}$, that is, every reachable state in $P||S^{'}$ can reach some state in $Q_{b,m}$. 
\end{itemize}

\emph{Proof}: Based on the definition of covertness and damage-nonblocking sensor attack, $A$ is a covert damage-nonblocking attack w.r.t. the attack constraint $(\Sigma_{o,a}, \Sigma_{s,a})$ if and only if any state in $Q_{bad}$ is not reachable in $P||A$ and every reachable state in $P||A$ can reach some state in $Q_{b,m}$.
Then, we can view the attack $A$ as a supervisor $S^{'}$ over the control constraint $\mathscr{C}_{ac}$ and $P$ as the new plant, which completes the proof. \hfill $\blacksquare$

By using the normality property, the following result follows straightforwardly from Theorem IV.1 and \cite{wonham2015supervisory}.

\emph{Corollary IV.1:} Given any $G_{new}$, $AC$, $OC$, $NS$, $CC$, and $M$, the supremal covert damage-nonblocking sensor attack $A$ w.r.t. the attack constraint $(\Sigma_{o,a}, \Sigma_{s,a})$ exists.

\emph{Proof}: Since the set of controllable events is a subset of the set of observable events in the attack-control constraint $\mathscr{C}_{ac}$, normality is equivalent to observability. In addition, since normality and controllability are closed under unions, the supremal covert damage-nonblocking sensor attack exists, which completes the proof. \hfill $\blacksquare$

Based on Theorem IV.1, the supremal damage-nonblocking sensor attack can be computed by adopting the normality based synthesis approach \cite{su2010model}, which is realized as \textbf{make\_supervisor}\footnote{In this algorithm, two kinds of states are pruned: 1) those ``bad'' states which break the covertness property; 2) those states which break the nonblockingness property.} in \textbf{SuSyNA} \cite{SuSyNA}. The requirement of the new plant $P$ can be generated by pruning those states in $P$ which break the covertness property. 
%{\color{red} I think Procedure 1 is not needed. }

%\emph{Procedure 1: (Compute the supremal damage-nonblocking sensor attack)}
%\begin{enumerate}[1)]
%    \item Generate $K = (Q_{k}, \Sigma_{k}, \xi_{k}, q_{k}^{init}, Q_{k,m})$, where
%    \begin{itemize}
%    \setlength{\itemsep}{3pt}
%    \setlength{\parsep}{0pt}
%    \setlength{\parskip}{0pt}
%        \item $Q_{k} := Q_{p} - Q_{del}$, $Q_{del} = \{((s,\gamma,q), q_{ac}, q_{oc}, q_{ns}, q_{cc},\\ q_{m}) \in Q_{p}|q \notin Q_{d} \wedge q_{m} = \varnothing\}$
%        \item $\Sigma_{k} = \Sigma_{p}$
%        \item $(\forall q_{1}, q_{2} \in Q_{k})(\forall \sigma \in \Sigma_{k}) \xi_{p}(q_{1}, \sigma) = q_{2} \Leftrightarrow \xi_{k}(q_{1}, \sigma) = q_{2}$  
%        \item $q_{k}^{init} = q_{p}^{init}$
%        \item $Q_{k,m} = Q_{p,m}$
%    \end{itemize}
%    \item Adopt \textbf{make\_supervisor} to compute the supremal damage-nonblocking sensor attack by treating $P$ as the plant and $K$ as the requirement.
%\end{enumerate}

%{\color{blue} I am wondering how about deleting those parts about damage-reachable attack. I am not sure how to formulate the computation of damage-reachable attack as Procedure 1.}

\emph{Theorem IV.2:} Given any $G_{new}$, $AC$, $OC$, $NS$, $CC$, and $M$, there exists a covert damage-reachable sensor attack w.r.t. the attack constraint $(\Sigma_{o,a}, \Sigma_{s,a})$ if and only if there exists a supervisor $S^{'}$ over attack-control constraint $\mathscr{C}_{ac} = (\Sigma_{s,a}^{\#} \cup \{stop\}, \Sigma_{o,a} \cup (\Sigma_{o,a} - \Sigma_{s,a})^{in} \cup \Sigma_{s,a}^{\#} \cup \{tick, stop\})$ such that 
\begin{itemize}
\setlength{\itemsep}{3pt}
\setlength{\parsep}{0pt}
\setlength{\parskip}{0pt}
    \item Any state in $Q_{bad}$ is not reachable in $P||S^{'}$.
    \item Some state in $Q_{b,m}$ is reachable in $P||S^{'}$, i.e., $L_{m}(P||S^{'}) \neq \varnothing$.
\end{itemize}

\emph{Proof}: Based on the definition of covertness and damage-reachable sensor attack, $A$ is a covert damage-reachable attack w.r.t. the attack constraint $(\Sigma_{o,a}, \Sigma_{s,a})$ if and only if any state in $Q_{bad}$ is not reachable in $P||A$ and $L_{m}(P||A) \neq \varnothing$.
Then, we can view the attack $A$ as a supervisor $S^{'}$ over the control constraint $\mathscr{C}_{ac}$ and $P$ as the new plant, which completes the proof. \hfill $\blacksquare$

%By using the normality property, the following result follows straightforwardly from the above transformation methodology and \cite{wonham2015supervisory}.

\emph{Corollary IV.2:} Given any $G_{new}$, $AC$, $OC$, $NS$, $CC$, and $M$, the supremal covert damage-reachable sensor attack $A$ w.r.t. the attack constraint $(\Sigma_{o,a}, \Sigma_{s,a})$ exists.

\emph{Proof}: The proof is similar to that of Corollary IV.1. \hfill $\blacksquare$

Based on Theorem IV.2, to compute the supremal damage-reachable sensor attack, slight changes on \textbf{make\_supervisor} of \textbf{SuSyNA} are needed, where now only those ``bad'' states breaking the covertness property are pruned while those states breaking the nonblockingness property are kept, since the definition of the damage-reachable sensor attack does not require that the closed behavior of $P||A$ is nonblocking.

%It is noteworthy that the control constraint for this new supervisor is $\mathscr{C}_{ac} = (\Sigma_{s,a}^{\#} \cup \{stop\}, \Sigma_{o,a} \cup (\Sigma_{o,a} - \Sigma_{s,a})^{in} \cup \Sigma_{s,a}^{\#} \cup \{tick, stop\})$, where the controllable events is a subset of the observable events.  

\subsection{Computational Complexity}
\label{subsec: Computational Complexity}
Next, we shall analyze the computational complexity of the proposed method. Firstly, the state size of the new plant $P$, denoted as $|Q_{p}|$, is at most 
\[
|Q_{cs}| \times |Q_{ce}| \times |Q| \times |Q_{ac}| \times |Q_{oc}| \times |Q_{ns}| \times |Q_{cc}| \times |Q_{m}|
\]
where
\begin{itemize}
\setlength{\itemsep}{3pt}
\setlength{\parsep}{0pt}
\setlength{\parskip}{0pt}
    \item $|Q_{cs}| \leq \frac{[|\Gamma|(\Delta_{s}+1)]^{C_{cs}+1}-1}{|\Gamma|(\Delta_{s}+1)-1}$
    \item $|Q_{ce}| \leq |\Gamma|(1 + \mathop {\max}\limits_{\sigma \in \Sigma_{c}}\{t_{e}^{\sigma}\})$
    \item $|Q_{ac}| = U + 2 + |\Sigma_{o} - \Sigma_{s,a}|$
    \item $|Q_{oc}| = \frac{[|\Sigma_{o}|(\Delta_{o}+1)]^{C_{oc}+1}-1}{|\Sigma_{o}|(\Delta_{o}+1)-1}$
    \item $|Q_{cc}| = \frac{[|\Gamma|(\Delta_{c}+1)]^{C_{cc}+1}-1}{|\Gamma|(\Delta_{c}+1)-1}$
    \item $|Q_{m}| \leq 2^{|Q_{cs}| \times |Q_{ce}| \times |Q| \times |Q_{oc}| \times |Q_{ns}| \times |Q_{cc}|}$
\end{itemize}
The alphabet of the new plant $P$, denoted as $\Sigma_{p}$, is 
\[
\Sigma \cup (\Sigma_{o} - \Sigma_{s,a})^{in} \cup \Sigma_{s,a}^{\#} \cup \Sigma_{o}^{out} \cup \Gamma^{in} \cup \Gamma^{out} \cup \Gamma \cup \{tick, stop\}
\]
%Then, the computational complexity depends on the chosen synthesis algorithm. 
The normality based synthesis approach~\cite{wonham2015supervisory},~\cite{su2010model},~\cite{WangStateControl2018} can be used for the synthesis of the supremal damage-nonblocking attack and the supremal damage-reachable attack.
The complexity for the synthesis of the supremal damage-nonblocking sensor attack is no more than $O(|\Sigma_P||Q_{p}|^{2}4^{|Q_{p}|})$, while the complexity for the synthesis of the supremal damage-reachable sensor attack is no more than $O(|\Sigma_P|2^{|Q_{p}|})$ \cite{WangStateControl2018}.

\section{Example}
\label{sec:example}
In this section, we shall present an example to show the effectiveness of the proposed method to synthesize covert sensor attacks for networked DES.

\begin{figure}[htp]
\begin{center}
\includegraphics[height=2.3cm]{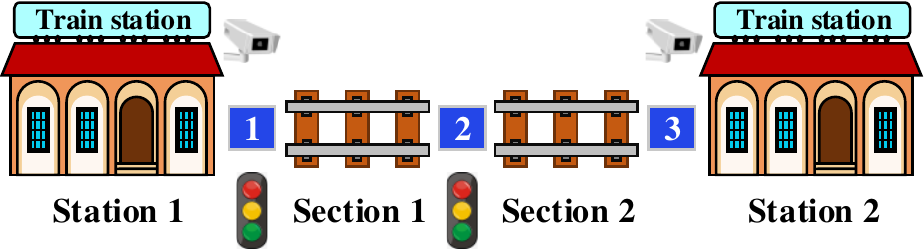}  
\caption{Illustration of guideway example}
\label{fig:Railway_Cartoon_Picture}
\end{center}                                
\end{figure}

\emph{Example V.1:} We adapt the mini-guideway example of~\cite{Lin:14} for an illustration. In Fig. \ref{fig:Railway_Cartoon_Picture}, there are two stations and a one-way track from station 1 to station 2. Two trains need to travel from station 1 to station 2. The track consists of 2 sections, with traffic lights and cameras installed at part of three junctions. Traffic lights control whether one train could pass the junction, and cameras detect whether one train has passed the junction. Two trains need to avoid the simultaneous occupation of the same section.

In this example, $\Sigma = \{a_{1},a_{2},a_{3},b_{1},b_{2},b_{3}\}$, where $a_{i}$ (respectively, $b_{i}$) means that train 1 (respectively, train 2) passes the junction $i$. $\Sigma_{uo} = \{a_{2},b_{2}\}$. $\Sigma_{uc} = \{a_{3},b_{3}\}$. $\Sigma_{s,a} = \Sigma_{o,a} = \Sigma_{o} = \{a_{1},a_{3},b_{1},b_{3}\}$. 
$\Sigma_{s,a}^{\#} = \{a_{1}^{\#},a_{3}^{\#},b_{1}^{\#},b_{3}^{\#}\}$.
$\Sigma_{o}^{out} = \{a_{1}^{out},a_{3}^{out},b_{1}^{out},b_{3}^{out}\}$. 
The set of control commands is $\Gamma = \{v_{1}, v_{2}, v_{3}\}$, where $v_{1}$ only allows train 1 to pass the junction, $v_{2}$ only allows train 2 to pass the junction, $v_{3}$ allows either train 1 or train 2 to pass the junction. $\Gamma^{in} = \{v_{1}^{in}, v_{2}^{in}, v_{3}^{in}\}$.
$\Gamma^{out} = \{v_{1}^{out}, v_{2}^{out}, v_{3}^{out}\}$.
$\Delta_{o} = 1$. $\Delta_{c} = 0$. $\Delta_{s} = 0$. $N_{f} = 1$. $U = 1$. $V = 1$. 

\begin{figure}[htbp]
\centering
\subfigure[]{
\begin{minipage}[t]{0.3\linewidth}
\centering
\includegraphics[height=1.3in]{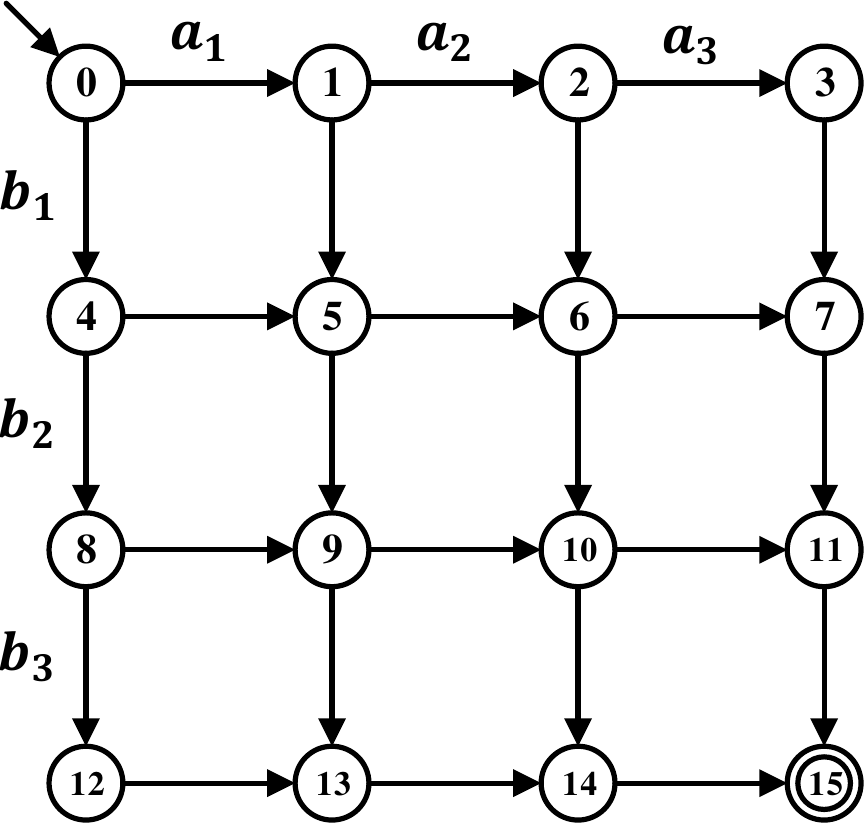}
%\caption{fig2}
\end{minipage}
}%
\subfigure[]{
\begin{minipage}[t]{0.8\linewidth}
\centering
\includegraphics[height=1.5in]{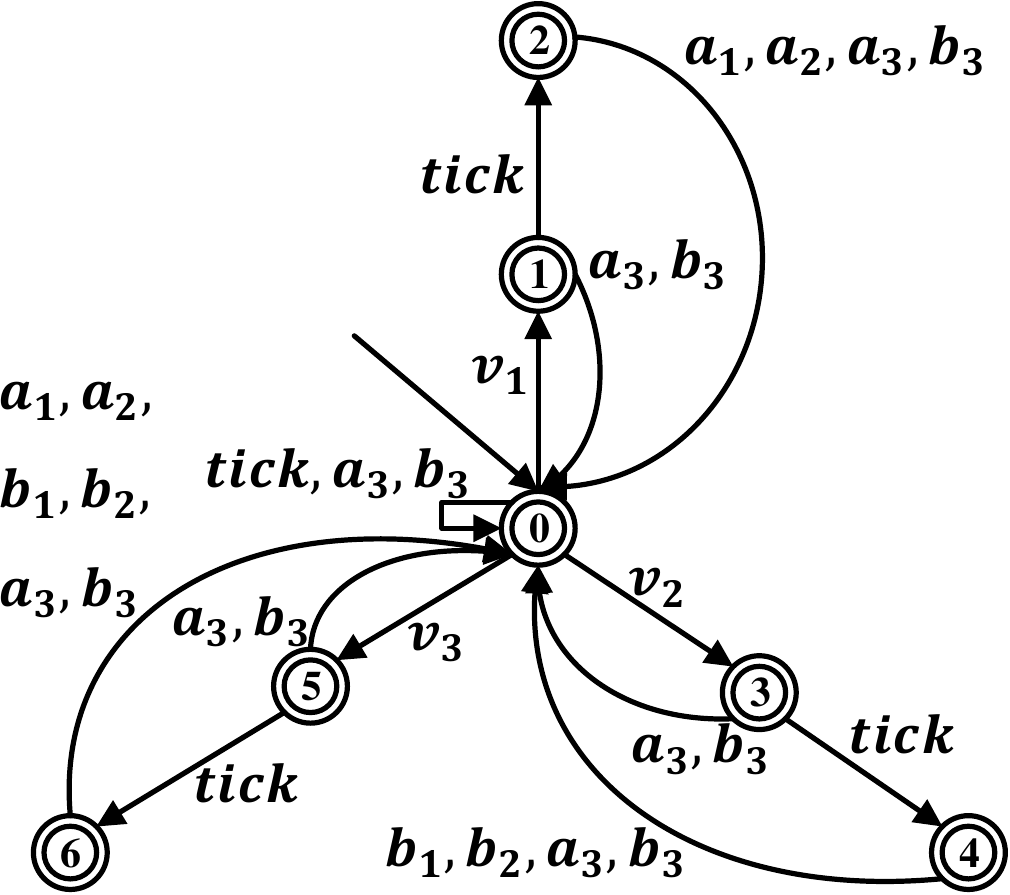}
%\caption{fig2}
\end{minipage}
}%

\centering
%\vspace{-0.2cm}
\caption{(a) Plant $G$. (b) Command execution $CE$.}
\label{fig:Example_G_CE}
\end{figure}

\begin{figure}[htbp]
\begin{center}
\includegraphics[height=2.5cm]{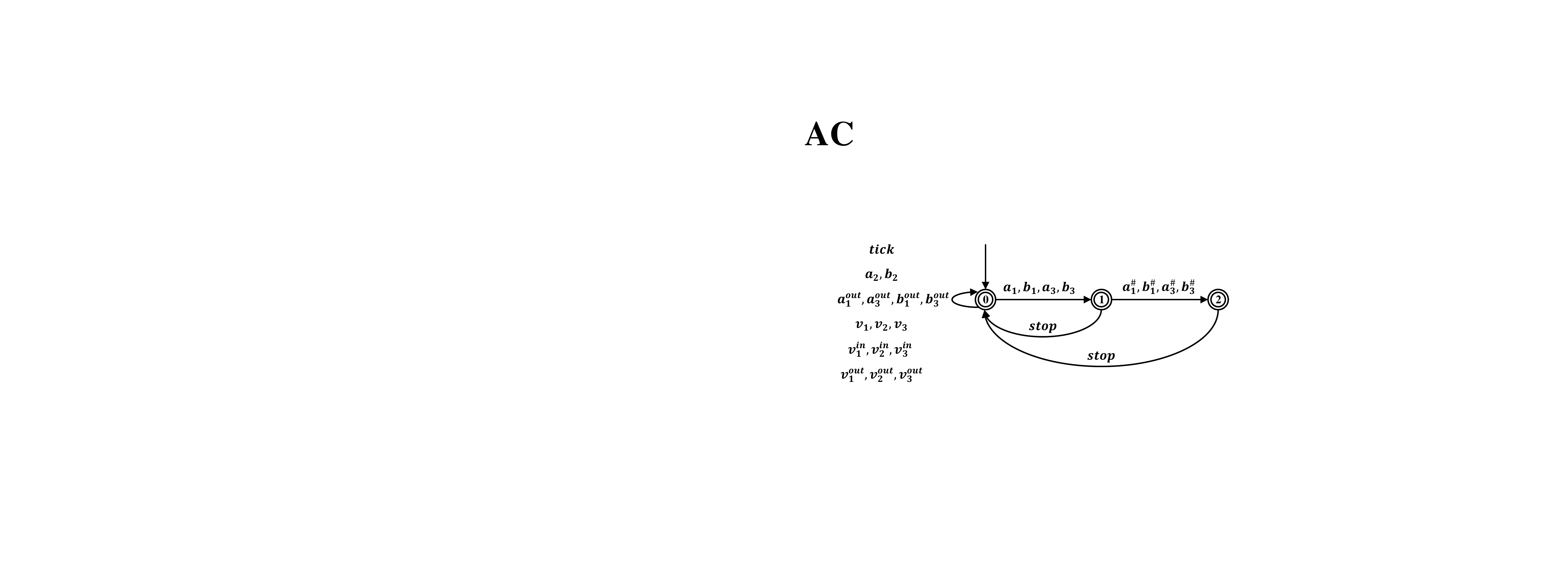}
\caption{Sensor attack constraints $AC$}  
\label{fig:Example_AC}
\end{center}                                 
\end{figure}

\begin{figure*}
\begin{center}
\includegraphics[height=10.3cm]{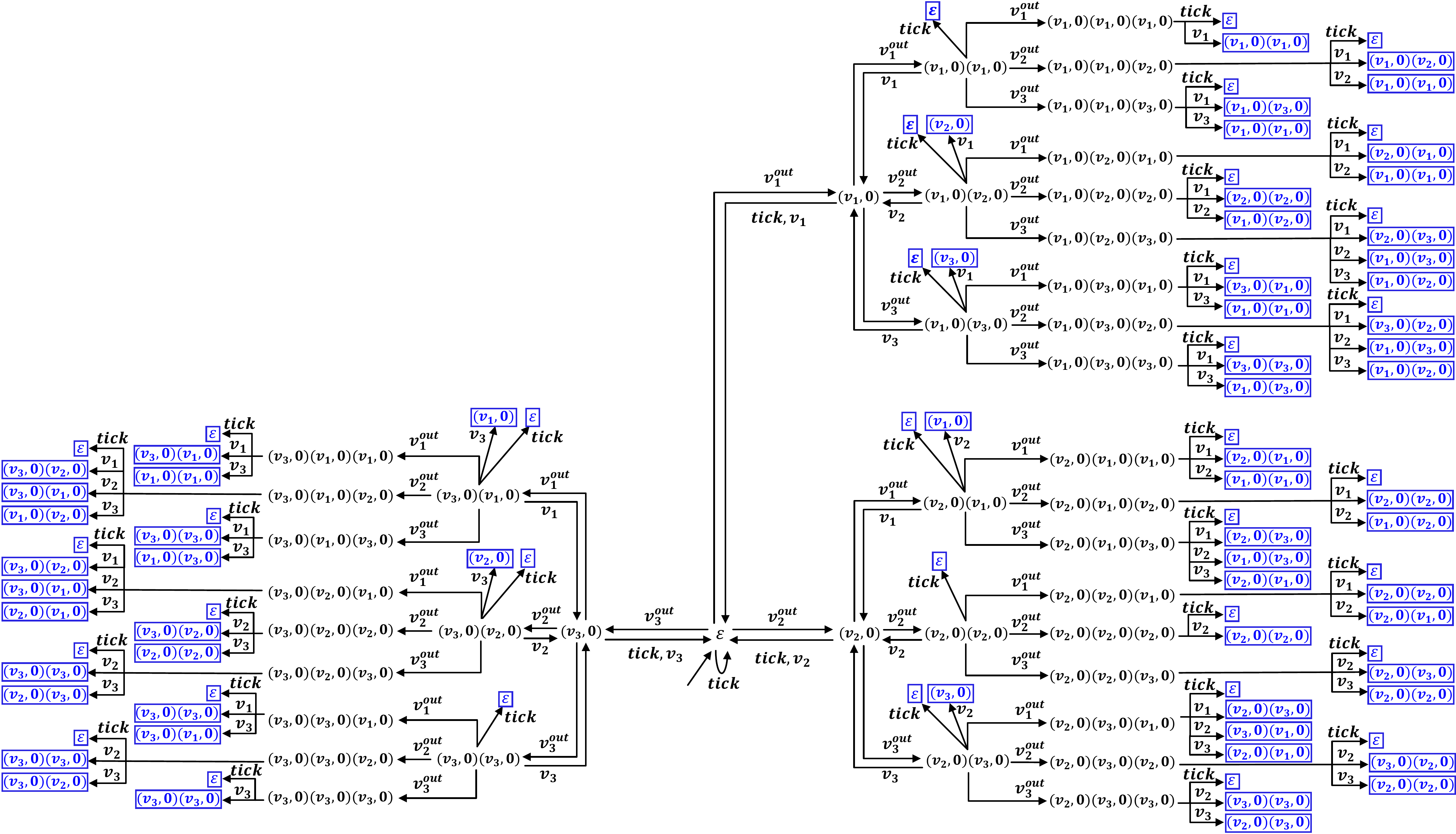}
\caption{Command storage $CS$}  
\label{fig:Example_CS}
\end{center}                                 
\end{figure*}

\begin{figure*}
\begin{center}
\includegraphics[height=6.8cm]{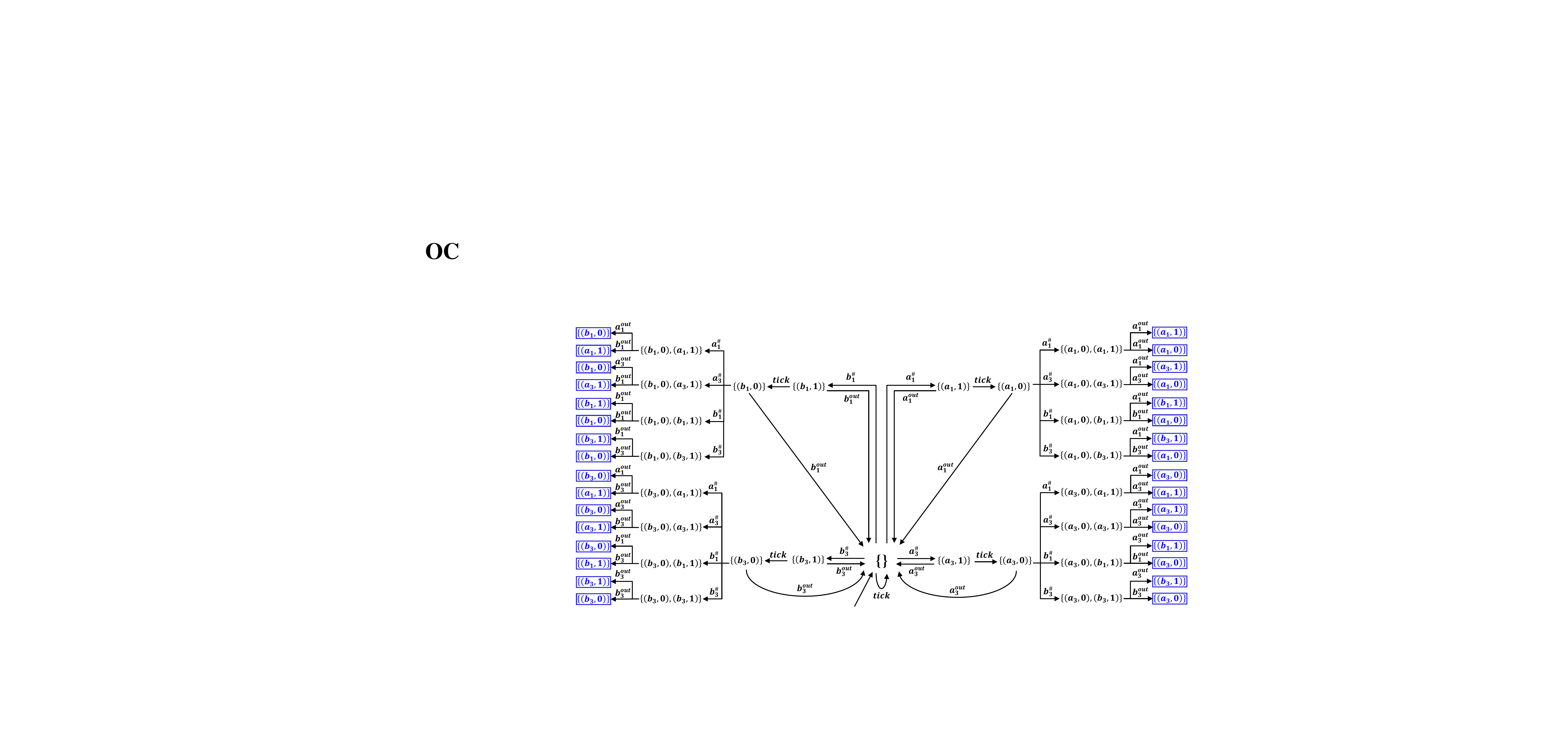}
\caption{Observation channel $OC$}  
\label{fig:Example_OC}
\end{center}                                 
\end{figure*}

\begin{figure*}
\begin{center}
\includegraphics[height=3.5cm]{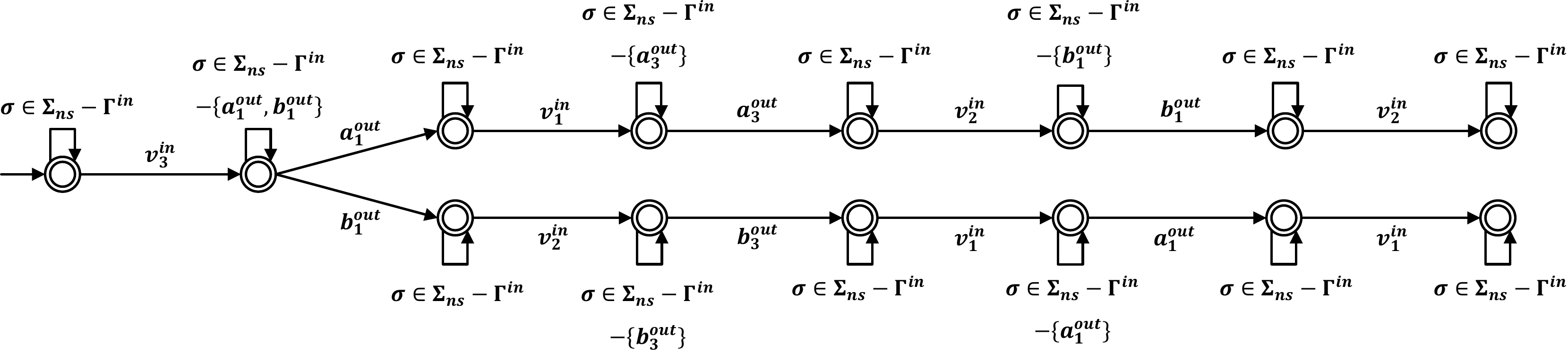}
\caption{Networked supervisor $NS$}  
\label{fig:Example_NS}
\end{center}                                 
\end{figure*}

\begin{figure*}
\begin{center}
\includegraphics[height=6.8cm]{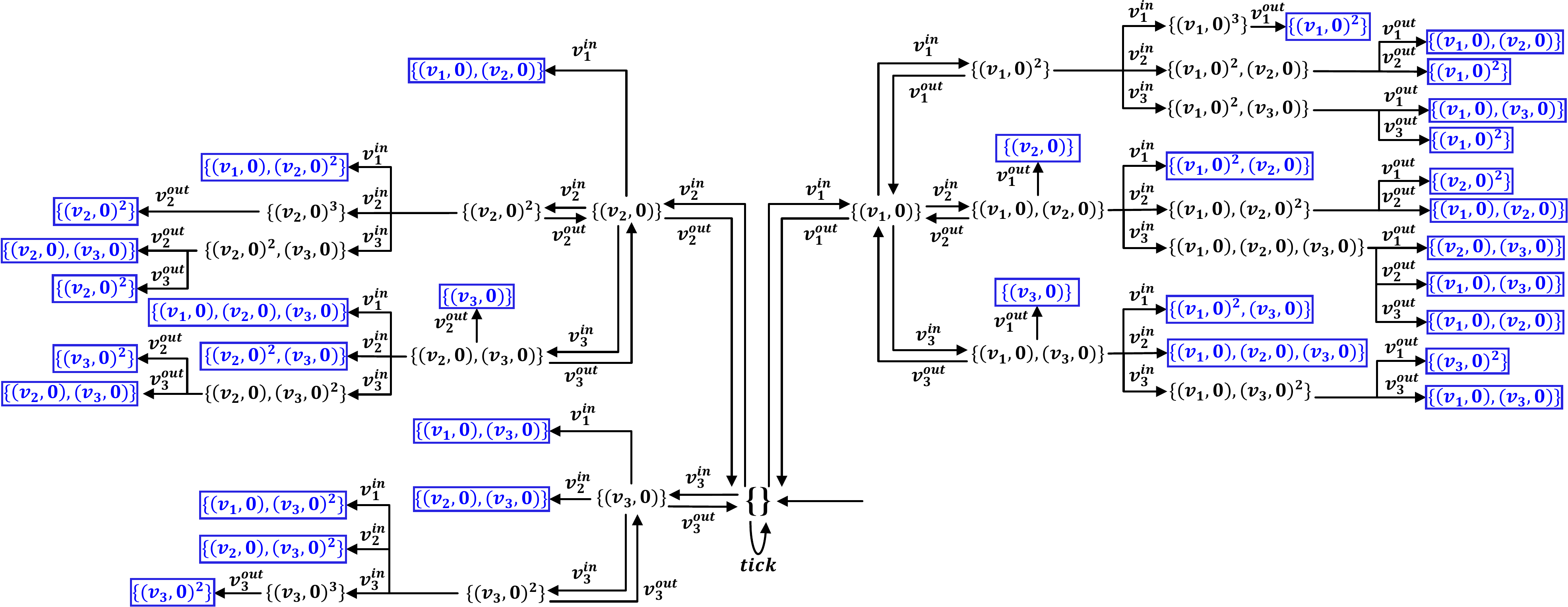}
\caption{Control channel $CC$}  
\label{fig:Example_CC}
\end{center}                                 
\end{figure*}

Plant $G$ is shown in Fig. \ref{fig:Example_G_CE}.(a), where the damage state set is $Q_{d} = \{5, 10\}$. 
Command execution $CE$ is shown in Fig. \ref{fig:Example_G_CE}.(b).
Sensor attack constraints $AC$ is shown in Fig. \ref{fig:Example_AC}.
Command storage $CS$ is shown in Fig. \ref{fig:Example_CS}.
Observation channel $OC$ is shown in Fig. \ref{fig:Example_OC}.
The designed networked supervisor\footnote{One possible way to synthesize the networked supervisor is: Firstly, based on the parameter $V$, we could model the networked supervisor constraints $NSC$ as a finite state automaton, which is similar to the model of sensor attack constraints. Then, since the preemption is not imposed in our framework, similar to the approach used in \cite{zhusupervisor,Linnetwork}, we could view $G_{new}||OC^{T}||NSC||CC$ as the new plant, where the difference is the introduced tick event, which is uncontrollable but observable to the networked supervisor. The specification for this new plant could be generated based on $G_{new}||OC^{T}||NSC||CC$. The networked supervisor could be viewed as the supervisor for such new plant w.r.t. the generated specification. Finally, the tool \textbf{SuSyNA} can be adopted to synthesize the networked supervisor.} $NS$ is shown in Fig. \ref{fig:Example_NS}. 
Control channel $CC$ is shown in Fig. \ref{fig:Example_CC}.
It can be checked that the designed $NS$ could fulfill the closed-behavior $\overline{\{a_{1}a_{2}a_{3}b_{1}b_{2}b_{3}, b_{1}b_{2}b_{3}a_{1}a_{2}a_{3}\}}$ of plant $G$ in the absence of sensor attacks. In addition, this networked supervisor satisfy the conditions defined in Section \ref{subsec:Networked Supervisor}: network controllability and network observability.
%{\color{blue} may need to say something more on the networked supervisor, why it is a valid networked supervisor, i.e., preserve controllability, observability for the networked setup? How it is synthesized. While it is straightforward for non-networked setup, it may not be the case for networked setup.}.
To make the drawn automata concise, in Fig. \ref{fig:Example_CS}, \ref{fig:Example_OC}, and \ref{fig:Example_CC}, some states are marked in blue rectangulars to denote that these states are repeated states. 
%{\color{red} not good expression, no idea how to say this} {\color{blue} is there a minimal automaton for each figure in the cfg file. how many states there, using color is not good as printed paper cannot distinguish colors.} 

\vspace{-0.3cm}

\begin{figure*}
\begin{center}
\includegraphics[height=7.7cm]{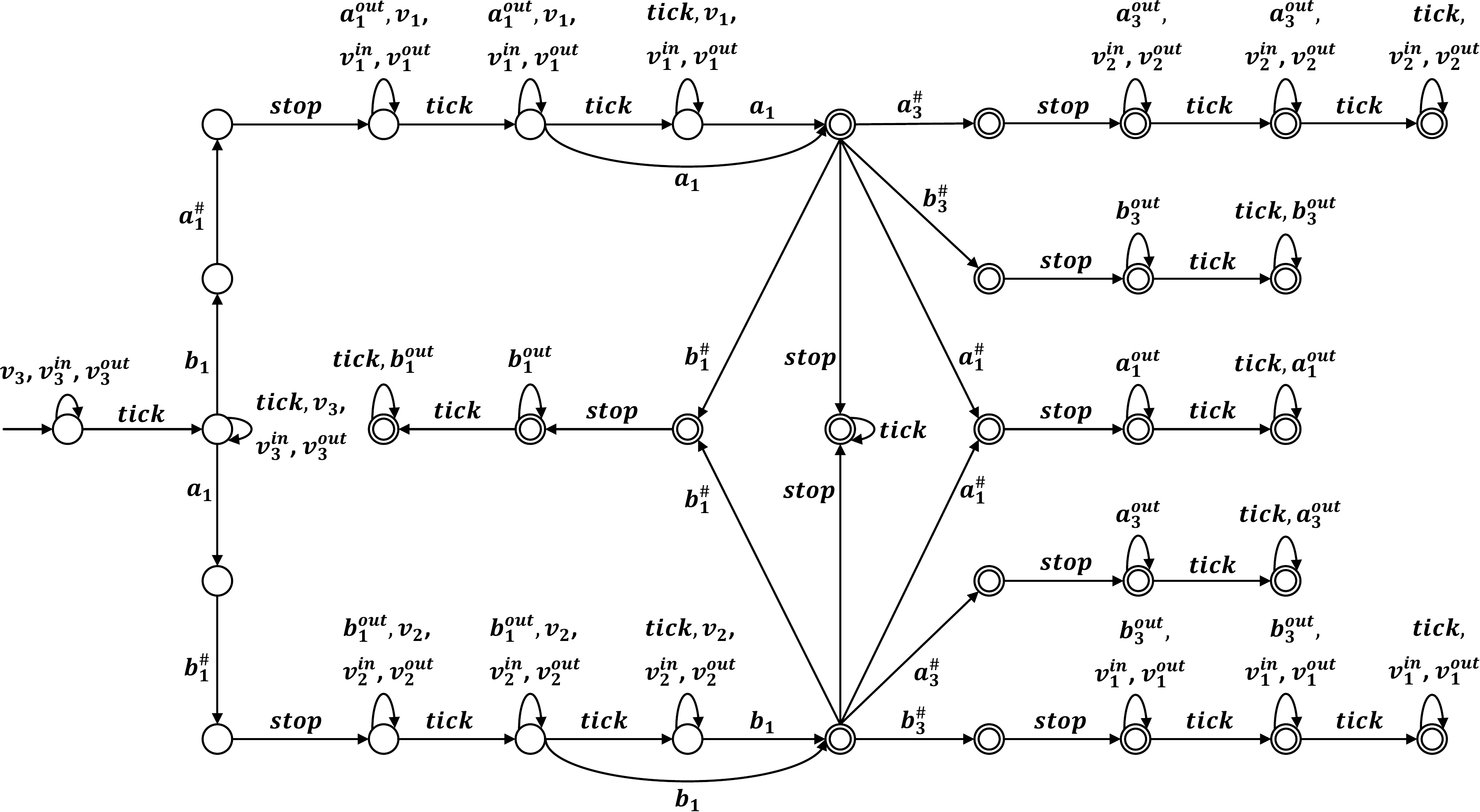}
\caption{Synthesized sensor attack $A$}  
\label{fig:Example_A}
\end{center}                                 
\end{figure*}

%To synthesize the desired sensor attack, we adopt \textbf{SuSyNA}, a tool which could deal with the non-deterministic plant automaton. 
%It is noteworthy that \textbf{TCT} can also be used for the synthesis of this example as explained in footnote 6 on page 6.

%{\color{red} better explain  TCT can also be used for the synthesis, if  footnote 6 in page 6 is kept there.} 
%The specific synthesis operations are as follows: {\color{blue} maybe no need to give more details here to confuse the reader. Readers familiar with SCT should be able to figure out the details.}
%\begin{enumerate}[1.]
%\setlength{\itemsep}{3pt}
%\setlength{\parsep}{0pt}
%\setlength{\parskip}{0pt}
%    \item Use \textbf{make\_product.py} to generate $G_{new}$ and $M$ {\color{blue} need further operations such as pruning, projection, self-looping}, based on which generate the new transformed plant $P = G_{new}||AC||OC||NS||CC||M$.
%    \item Generate the specification of the new transformed plant by pruning those states which belong to $Q_{bad}$ since the attack needs to be covert.
%    \item Use \textbf{make\_supervisor.py} to synthesize the supervisor with $P$ as the plant and the specification generated in Step 2 as the requirement.
%    \item Use \textbf{make\_feasible\_supervisor.py} and  \textbf{make\_\\minimization.py} to transform the supervisor synthesized in Step 3 to an equivalent minimized feasible supervisor, which is the target sensor attack. 
%\end{enumerate}
For brevity, we here shall only synthesize the damage-nonblocking sensor attack, which is shown in Fig. \ref{fig:Example_A}. The strategy of this sensor attack is: At initial state, the attacker will wait for at least one $tick$, then the first time the attacker observes $a_{1}$ (or $b_{1}$), it will implement replacement attacks, that is, alter the sensor reading and send $b_{1}$ (or $a_{1}$) into the observation channel. By this way, the networked monitor will not detect the existence of the attack meanwhile the networked supervisor will issue the wrong control command $v_{2}$ (or $v_{1}$). After receiving the control command, plant $G$ will execute $b_{1}$ (or $a_{1}$) and transit to the damage state 5 in the plant $G$, i.e., the attack goal is achieved. Before observing a sensor reading again, the attacker will wait for at least one $tick$; afterwards, no matter what the attacker observes, it could implement either deletion attack or replacement attack. For example, if the attacker observes $a_{1}$, then it could delete $a_{1}$ and send nothing into the observation channel or replace the sensor reading by anyone of $a_{1}$, $a_{3}$, $b_{1}$, and $b_{3}$. The only difference among these four replacement attack strategies is that sending $a_{3}$ will still enable the attack to remain covert while sending anyone of $a_{1}$, $b_{1}$, and $b_{3}$ will expose its existence. This exposure is allowed since the plant $G$ has been induced to the damage state.

\section{Conclusions}
\label{sec:conclusions}
In this paper, we propose an approach to synthesize the covert sensor attacks in networked DES with non-FIFO channels by modeling the original problem as the Ramadge-Wonham supervisory synthesis problem under partial observation for both the damage-reachable goal and damage-nonblocking goal, allowing existing synthesis tools such as SuSyNA or TCT to be used. This makes the proposed framework attractive to practitioners, who may quickly adopt the framework without any further major efforts on synthesis tool development.    
Based on the constructed model, the supremal covert sensor attack exists for each type of sensor attack, and is computable, although the computational complexity is not the major concern of this paper, and shall be addressed in our follow-up works on structural synthesis methods, e.g., a distributed synthesis method.  
In our future works, we shall 
%extend the framework to investigate the covert sensor-actuator attacks in networked DES to exploit more attack strategies and 
also explore synthesis of resilient supervisors against such covert attacks in networked DES.
%; on the other hand, it is of great research value to investigate the resilient supervisor synthesis in networked DES to guarantee the safe operations of CPS.

%\begin{ack}                               % Place acknowledgements
%The funding supports from Singapore National Research Foundation via Delta-NTU Corporate Lab Program (DELTA-NTU CORP
%LAB-SMA-RP2 SU RONG M4061925.043) and from Singapore Ministry of Education Tier
%1 Academic Research Grant (M4011982 RG91/18-(S)-SU RONG (VP)) are gratefully
%acknowledged.  % here.
%\end{ack}

%\vspace{0.4cm}
\begin{appendices} 
\section{Proof of Theorem 3.1} 
\label{appendix: 1}
Based on the basic setup in this work: 1) the largest number of fired events without ticks at plant $G$ is $N_{f}$; 2) the maximum number of events that the attacker can simultaneously send into the observation channel is $U$; 3) the upper bound of delay in the observation channel is $\Delta_{o}$, it can be derived that  
\[
N_{o}(t) \leq N_{f}U(\Delta_{o} + 1)
\]
In addition, $N_{f}U(\Delta_{o} + 1)$ can be achieved in the following case. We consider consecutive $(\Delta_{o} + 1)$ discrete time: $t$, $t+1$, $\dots$, $t+\Delta_{o}$. At each time, the following three conditions are satisfied: 1) Plant $G$ fire $N_{f}$ events in $\Sigma_{o,a}$; 2) the attacker always sends $U$ events after observing one event; 3) any event sent into the observation channel is delayed for $\Delta_{o}$ ticks. Thus, at time $t+\Delta_{o}$, the number of events in the observation channel is $N_{f}U(\Delta_{o} + 1)$. This completes the proof. \hfill $\blacksquare$
%\vspace{0.4cm}
\section{Proof of Theorem 3.2} 
\label{appendix: 2}
Firstly, we prove that $N_{c}(t) \leq N_{f}UV(\Delta_{o} + \Delta_{c} + 1) + V(\Delta_{c} + 1)$.
Since the upper bound of delay in the control channel is $\Delta_{c}$, we only need to consider control commands entering the control channel during consecutive $(\Delta_{c} + 1)$ discrete time. Otherwise, if we consider $N > \Delta_{c} + 1$ consecutive discrete time, then the control commands entering the observation channel during the first $N-(\Delta_{c} + 1)$ time must have been popped out from the channel at the latest time of these $N$ consecutive time, which means that the problem is reduced to the case of consecutive $(\Delta_{c} + 1)$ discrete time.
%{\color{blue} intuitively looks true, but things can be tricky,  need to check later on this statement.} 
Without loss of generality, we choose time $t + \Delta_{o}$, $t + \Delta_{o} + 1$, $\dots$, $t + \Delta_{o} + \Delta_{c}$.
In addition, it is straightforward to derive that the maximum number of events in the observation channel is achieved only when the following three conditions are satisfied: At each time, 1) Plant $G$ fires $N_{f}$ events in $\Sigma_{o,a}$; 2) the attacker sends $U$ events after observing one event; 3) networked supervisor issues $V$ control commands after observing $tick$ or an observable event popped out of the observation channel. 
%{\color{blue} may need to unify with the conditions in the first paragraph of the left column of this page, which also gives 3) conditions, into four conditions? better to be the same..not explaining part of the conditions in each case.}. 

Based on the above analysis, at any time, the number of events entering the observation channel is $N_{f}U$. The delay of the $i$-th event of these $N_{f}U$ events entering the observation channel at time $j$ is denoted as $d^{oc}(j, i)$.
Then, at time $t^{'}$, the number of events popped out from the observation channel, i.e., observed by the networked supervisor, is 
\[
\sum\limits_{j = t^{'}-\Delta_{o}}^{t^{'}}\sum\limits_{i = 1}^{N_{f}U}\mathbb{B}^{oc}(t^{'}, j, i)
\]
where $\mathbb{B}^{oc}: \mathbb{N} \times \mathbb{N} \times \mathbb{N}^{+} \rightarrow \{0,1\}$ is a mapping defined as 
\[
\mathbb{B}^{oc}(t^{'}, j, i)=
\left\{
\begin{array}{rcl}
1       &      & {d^{oc}(j,i) = t^{'}-j, i \in \mathbb{N}^{+}}\\
0  &      & {\rm otherwise.}
\end{array} \right.
\]
it describes whether the $i$-th event entering into the observation channel at time $j$ 
%{\color{blue} by definition (see the first sentence of this paragraph, following the last sentence of the last page), these $N_fU$ messages all enter the channel at the same time instant $t'$? The proof idea looks interesting.} 
with delay $d^{oc}(j,i)$ will be popped out from the channel at time $t^{'}$. If $d^{oc}(j,i) = t^{'}-j$, then it will be popped out; otherwise, it will not.
Thus, at time $t^{'}$, the number of control commands entering the control channel is 
\[
V\sum\limits_{j = t^{'}-\Delta_{o}}^{t^{'}}\sum\limits_{i = 1}^{N_{f}U}\mathbb{B}^{oc}(t^{'}, j, i) + V
\]
since $tick$ could also trigger the networked supervisor to issue control commands. Then the total number of control commands entering the control channel during $[t + \Delta_{o}: t + \Delta_{o} + \Delta_{c}]$ is 
\[
\begin{aligned}
&V\sum\limits_{j = t}^{t + \Delta_{o}}\sum\limits_{i = 1}^{N_{f}U}\mathbb{B}^{oc}(t + \Delta_{o}, j, i) + V + \dots \\ &+ V\sum\limits_{j = t + \Delta_{c}}^{t + \Delta_{o} + \Delta_{c}}\sum\limits_{i = 1}^{N_{f}U}\mathbb{B}^{oc}(t + \Delta_{o} + \Delta_{c}, j, i) + V
\\ & = V[\sum\limits_{j = t}^{t + \Delta_{o}}\sum\limits_{i = 1}^{N_{f}U}\mathbb{B}^{oc}(t + \Delta_{o}, j, i) + \dots \\ &+ \sum\limits_{j = t + \Delta_{c}}^{t + \Delta_{o} + \Delta_{c}}\sum\limits_{i = 1}^{N_{f}U}\mathbb{B}^{oc}(t + \Delta_{o} + \Delta_{c}, j, i)] + V(\Delta_{c}+1)
\\ & \leq N_{f}UV(\Delta_{o} + \Delta_{c} + 1) + V(\Delta_{c}+1)
\end{aligned}
\]
since given $i$ and $j$, the value of $t^{'}$ such that $\mathbb{B}^{oc}(t^{'}, j, i) = 1$ is unique.

Next, we show that this maximum can be achieved. Based on Theorem III.1, at time $t+\Delta_{o}$, assuming that there are $N_{f}U(\Delta_{o}+1)$ events in the observation channel and all of these events need to be popped out now. Then at time $t+\Delta_{o}$, networked supervisor issues $N_{f}UV(\Delta_{o}+1) + V$ control commands into the control channel. In the following $\Delta_{c}$ times, $G$ fires $N_{f}$ events in $\Sigma_{o,a}$ at each time. The attacker sends $N_{f}U$ events sent into the observation channel without delays at each time. The supervisor issues $(N_{f}U+1)V$ control commands into the control channel at each time. For the control commands entering the control channel at time $t+\Delta_{o}$, $\dots$, $t+\Delta_{o}+\Delta_{c}$, assuming that each control command is delayed for $\Delta_{c}$ ticks, then, at time $t+\Delta_{o}+\Delta_{c}$, the number of control commands in the control channel is 
\[
\begin{aligned}
&N_{f}UV(\Delta_{o}+1) + V + \Delta_{c}(N_{f}U+1)V \\ &= N_{f}UV(\Delta_{o} + \Delta_{c} + 1) + V(\Delta_{c}+1) 
\end{aligned}
\]
This completes the proof. \hfill $\blacksquare$
%\vspace{-0.3cm}
\section{Proof of Theorem 3.3} 
\label{appendix: 3}
Firstly, we prove that $N_{cs}(t) \leq N_{f}UV(\Delta_{o} + \Delta_{c} + \Delta_{s} + 1) + V(\Delta_{c} + \Delta_{s} + 1)$. Since the upper bound of storage time is $\Delta_{s}$, we only need to consider control commands entering the command storage module during consecutive $(\Delta_{s} + 1)$ discrete time. Without loss of generality, we choose time $t+\Delta_{o}+\Delta_{c}$, $\dots$, $t+\Delta_{o}+\Delta_{c}+\Delta_{s}$. In addition, the maximum number of control commands in the command storage module is achieved only when the following three conditions are satisfied: At each time, 1) Plant $G$ fires $N_{f}$ events in $\Sigma_{o,a}$; 2) the attacker sends $U$ compromised events after observing one event; 3) networked supervisor issues $V$ control commands after observing $tick$ or an observable event popped out of the observation channel. Then, at any time $t^{'} \in [t+\Delta_{o}+\Delta_{c}: t+\Delta_{o}+\Delta_{c}+\Delta_{s}]$, the number of control commands entering the control channel is denoted as
\[
n_{cc}^{e}(t^{'}) = V\sum\limits_{j = t^{'}-\Delta_{o}}^{t^{'}}\sum\limits_{i = 1}^{N_{f}U}\mathbb{B}^{oc}(t^{'}, j, i) + V
\]
The delay of $h$-th control command entering the control channel at time $k$ is denoted as $d^{cc}(k,h)$.
Then the number of control commands popped out from the control channel at time $t^{'}$, i.e., received by the command storage module, is
\[
\sum\limits_{k = t^{'}-\Delta_{c}}^{t^{'}}\sum\limits_{h = 1}^{n_{cc}^{e}(k)}\mathbb{B}^{cc}(t^{'}, k, h)
\]
where $\mathbb{B}^{cc}: \mathbb{N} \times \mathbb{N} \times \mathbb{N}^{+} \rightarrow \{0,1\}$ is a mapping defined as 
\[
\mathbb{B}^{cc}(t^{'}, k, h)=
\left\{
\begin{array}{rcl}
1       &      & {d^{cc}(k,h) = t^{'}-k, h \in \mathbb{N}^{+}}\\
0  &      & {\rm otherwise.}
\end{array} \right.
\]
it describes whether the $h$-th control command entering the control channel at time $k$ with delay $d^{cc}(k,h)$ will be popped out from the channel at time $t^{'}$.
Thus, the total number of control commands entering the command storage module during $[t+\Delta_{o}+\Delta_{c}: t+\Delta_{o}+\Delta_{c}+\Delta_{s}]$ is
\[
\begin{aligned}
&\sum\limits_{k = t+\Delta_{o}}^{t+\Delta_{o}+\Delta_{c}}\sum\limits_{h = 1}^{n_{cc}^{e}(k)}\mathbb{B}^{cc}(t+\Delta_{o}+\Delta_{c}, k, h) + \dots \\ &+ \sum\limits_{k = t+\Delta_{o}+\Delta_{s}}^{t+\Delta_{o}+\Delta_{c}+\Delta_{s}}\sum\limits_{h = 1}^{n_{cc}^{e}(k)}\mathbb{B}^{cc}(t+\Delta_{o}+\Delta_{c}+\Delta_{s}, k, h)
\\ &\leq 
\sum\limits_{k = t+\Delta_{o}}^{t+\Delta_{o}+\Delta_{c}}\sum\limits_{h = 1}^{n_{cc}^{e}(k)-V}\mathbb{B}^{cc}(t+\Delta_{o}+\Delta_{c}, k, h) + \dots \\ &+ \sum\limits_{k = t+\Delta_{o}+\Delta_{s}}^{t+\Delta_{o}+\Delta_{c}+\Delta_{s}}\sum\limits_{h = 1}^{n_{cc}^{e}(k)-V}\mathbb{B}^{cc}(t+\Delta_{o}+\Delta_{c}+\Delta_{s}, k, h) + \\ & V(\Delta_{c} + \Delta_{s} + 1)
\\ & \leq N_{f}UV(\Delta_{o} + \Delta_{c} + \Delta_{s} + 1) + V(\Delta_{c} + \Delta_{s} + 1)
\end{aligned}
\]
since given $i$ and $j$, the value of $t$ such that $\mathbb{B}^{oc}(t^{'}, j, i) = 1$ is unique, and given $h$ and $k$, the value of $t^{'}$ such that $\mathbb{B}^{cc}(t^{'}, k, h) = 1$ is unique.

Next, we show that this maximum can be achieved. Based on Theorem III.2, at time $t+\Delta_{o}+\Delta_{c}$, assuming that there are $N_{f}UV(\Delta_{o} + \Delta_{c} + 1) + V(\Delta_{c} + 1)$ in the control channel and all of them need to be popped out. Thus, at time $t+\Delta_{o}+\Delta_{c}$, there are $N_{f}UV(\Delta_{o} + \Delta_{c} + 1) + V(\Delta_{c} + 1)$ control commands stored in the command storage module. At each time of the following $\Delta_{s}$ times, 1) $G$ always fires $N_{f}$ events in $\Sigma_{o,a}$; 2) there are $N_{f}U$ events sent into the observation channel without delays; 3) networked supervisor issues $(N_{f}U+1)V$ control commands into the control channel without delays, then, at time $t+\Delta_{o}+\Delta_{c}+\Delta_{s}$, the number of stored control commands in the command storage is
\[
\begin{aligned}
&N_{f}UV(\Delta_{o} + \Delta_{c} + 1) + V(\Delta_{c} + 1) + \Delta_{s}(N_{f}U+1)V 
\\ &= N_{f}UV(\Delta_{o} + \Delta_{c} + \Delta_{s} + 1) + V(\Delta_{c} + \Delta_{s} + 1)
\end{aligned}
\]
This completes the proof. \hfill $\blacksquare$
\end{appendices}

\end{document}